\documentclass{aa}  

\usepackage{graphicx}
\usepackage{subcaption}
\usepackage{booktabs}
\usepackage[hidelinks]{hyperref}
\usepackage{xcolor}
\hypersetup{
    colorlinks,
    linkcolor={red!50!black},
    citecolor={blue!50!black},
    urlcolor={blue!80!black}
}
\usepackage[normalem]{ulem}
\usepackage{txfonts}
\begin{document}

   \title{Is XRISM/Resolve probing a "raining" absorber  in Mrk~509?}

   \author{M. Dadina\inst{1}
       \and V. Missaglia\inst{1}
        \and V. Braito\inst{2}
        \and M. Cappi\inst{1}
        \and A. Luminari\inst{3,4}
        \and D. Barret\inst{5}
        \and E. Bertola\inst{6}
        \and S. Bianchi\inst{7}
        \and A. Comastri\inst{1}
        \and G. Chartas\inst{8}
         \and J. Kaastra\inst{9,10}
        \and E. Kammoun\inst{11}
        \and G. Lanzuisi\inst{1}
        \and G. Matzeu\inst{12}
        \and R. Middei\inst{3,13}
         \and E. Nardini\inst{6}
        \and F. Nicastro\inst{3}
        \and Pierre-Olivier Petrucci\inst{14}
        \and C. Pinto\inst{15}
        \and R. Serafinelli\inst{16,3}
        \and A. Tortosa\inst{3}
        \and C. Vignali\inst{1,17}}

   \institute{$^1$ INAF – Osservatorio di Astrofisica e Scienza dello Spazio di Bologna, Via Gobetti 101, I-40129 Bologna, Italy \\ 
   $^2$ INAF – Osservatorio Astronomico di Brera, Via Bianchi 46 I-23807 Merate (LC), Italy  \\ 
   $^3$ INAF – Osservatorio Astronomico di Roma, Via Frascati 33, I-00078 Monte Porzio Catone, Italy\\
   $^4$ INAF-IAPS – Istituto di Astrofisica e Planetologia Spaziali, Via del Fosso del Caveliere 100, I-00133 Roma, Italy\\
   $^5$ Institut de Recherche en Astrophysique et Plan\'etologie, 9 avenue du Colonel Roche, Toulouse, 31028, France\\
$^{6}$ INAF – Osservatorio Astrofisco di Arcetri, largo E. Fermi 5, 50127, Firenze, Italy\\
   $^7$ Dipartimento di Matematica e Fisica, Università degli Studi Roma Tre, Via della Vasca Navale 84, I-00146, Roma, Italy\\
   $^8$ Department of Physics and Astronomy, College of Charleston,
Charleston, SC 29424, USA \\
 $^9$  SRON Netherlands Institute for Space Research, Leiden, The Netherlands\\
 $^{10}$ Leiden Observatory, University of Leiden, P.O. Box 9513, NL-2300 RA, Leiden, The Netherlands\\ 
 $^{11}$ Cahill Center for Astronomy \& Astrophysics, California Institute of Technology, 1216 East California Boulevard, Pasadena, CA 91125, USA\\
 $^{12}$ Quasar Science Resources SL for ESA, European Space Astronomy Centre (ESAC), Science Operations Department, 28692, Villanueva
de la Ca\~{n}ada, Madrid, Spain\\
 $^{13}$ Space Science Data Center, Agenzia Spaziale Italiana, Via del Politecnico snc, 00133 Roma, Italy\\
 $^{14}$ Universit\'e Grenoble Alpes, CNRS, IPAG, 38000 Grenoble, France\\
 $^{15}$  INAF–IASF Palermo, Via U. La Malfa 153, I-90146 Palermo, Italy\\
 $^{16}$ Instituto de Estudios Astrofísicos, Facultad de Ingeniería y Ciencias, Universidad Diego Portales, Av. Ejército Libertador 441, Santiago, Chile\\
 $^{17}$ Dipartimento di Fisica e Astronomia ‘Augusto Righi’, Università degli Studi di Bologna, via Gobetti 93/2, I-40129 Bologna, Italy\\
             \email{mauro.dadina@inaf.it}
            }

   \date{\today}

 
  \abstract
   {X-ray spectroscopy of active galactic nuclei (AGN) offers unique insights into the reprocessing of radiation and the gas dynamics near supermassive black holes. The Seyfert 1 galaxy Mrk~509 is an ideal laboratory for these studies. It is in fact known for its complex Iron K emission and one of the four/five AGN to date in which evidence of transient, blue- and red-shifted absorption features indicative of high-velocity flows has been detected in the X-rays.}
   {We present the first high-resolution 2--10\,keV spectrum of Mrk~509 obtained with the \textit{XRISM}/Resolve calorimeter. Our primary goals are to disentangle the narrow and broad emission features and to place stringent constraints on the kinematics and physical location of circumnuclear gas flows, specifically searching for signatures of accretion and feedback.}
   {We analyzed 106 ks of XRISM/Resolve data, complemented by simultaneous \textit{XMM-Newton} and \textit{NuSTAR} observations, which we used to constrain the broadband continuum. We modeled the spectra using self-consistent reflection models for the continuum and the emission lines, and employed photoionized plasma models to characterize discrete, ionized absorption features.}
   { The XRISM/Resolve spectrum reveals a narrow Fe~K$\alpha$ core resolved with $\sigma\simeq10$\,eV (v$_{FWHM} \sim$ 1100 km/s). The best-fit model includes a broader component with $\sigma\sim450$\,eV. We also find tentative evidence for discrete absorption features at rest-frame energies of $E\sim6.2$\,keV. When modeled as an ionized absorber (with a significance of $\sim3.6\sigma$), the data suggest the gas is redshifted relative to the systemic velocity, corresponding to an inflow of material with $v_{\rm in}\sim11\,000$\,km\,s$^{-1}$, located within the inner few thousand gravitational radii.}
   {The narrow Fe~K$\alpha$ emission is consistent with an origin in the dusty torus, while the broad component arises from Broad Line Region scales or the accretion disk ($R\sim30$--$120\,r_g$). This scenario is strongly corroborated by our relativistic reflection modeling, which restricts the inner edge of the emitting region to distances greater than 27\,$r_g$. If confirmed by future observations, the high-velocity inflow would likely represent dense, fragmented clumps of a ``failed wind'' raining back onto the accretion disk, providing potential direct evidence that non-standard accretion processes may coexist with canonical disk-like flows in the inner regions of AGN.}


   \maketitle
%

\section{Introduction}

  X-ray spectroscopy of active galactic nuclei (AGN) provides an unparalleled tool for probing the extreme environment surrounding supermassive black holes (SMBHs), which are believed to co-evolve with their host galaxies \citep[e.g.,][]{kormendy,heckman}. Among the features observed, the iron K$\alpha$ (Fe K$\alpha$) emission line at 6.4 keV is nearly ubiquitous and serves as a powerful tracer of circumnuclear gas \citep{nandra07}. The line's profile is a key diagnostic of its physical origin: a narrow core is typically associated with reflection from distant material, such as the molecular torus \citep[see][]{antonucci1993}, while a broad and relativistically skewed profile is the hallmark of emission from the inner accretion disk, just a few gravitational radii ($r_g = \frac{GM}{c^2}$, where $G$ is the gravitational constant, $M$ is the mass of the SMBH, and $c$ is the speed of light) from the SMBH \citep[e.g.][]{tanaka, fabian2000}.

In addition to emission, AGN spectra often exhibit absorption features from ionized gas along the line of sight. Blue-shifted absorption lines are clear signatures of outflows, with the most extreme examples being the ultra-fast outflows (UFOs) that reach velocities up to v$\sim$0.1$-$0.5$c$ \citep[e.g.,][]{chartas, chartas21,pounds03,reeves02, reeves09, reeves14, reeves18, dadina, dadina18, cappi2006, cappi2009, tombesi10, tombesi, tombesi2013, gofford, nardini, matzeu16, matzeu17, matzeu23, braito, gianolli, pds456}. These powerful winds are considered a key channel for AGN feedback, regulating SMBH growth and star formation \citep[e.g.,][]{king, gaspari, laha}. Conversely, red-shifted absorption features, indicative of infalling material, remain far rarer. Only a small number of candidate ultra-fast inflows (UFIs) exist in the literature \citep[e.g.,][]{nandra99, dadina, yaqoob,longinotti,giustini,peca25}. UFIs offer a unique window into non-standard accretion processes, such as fallback from  ``aborted jets'' or  ``failed winds'', where the launched material fails to escape the SMBH's gravitational pull \citep[e.g.,][]{ghisellini,proga,giustini19}.

The bright, nearby Seyfert 1 galaxy Mrk 509 \citep[$z$=0.034,][]{huchra} is an ideal laboratory for studying these phenomena. Extensive monitoring has revealed a complex Fe K$\alpha$ structure composed of a constant narrow core and a variable, resolved component \citep[][]{ponti}. Reverberation mapping has shown that the resolved component responds to continuum variations on timescales of days, placing its origin within a few light-days from the central engine \citep[][]{ponti}. Furthermore, Mrk 509 is one of the very few objects where transient, red-shifted Fe K absorption lines—the potential signature of a UFI—have been reported in the literature \citep[][]{dadina}. The source has also shown evidence of variable, highly ionized outflows with velocities up to $\sim$ 14\,000 km s$^{-1}$ \citep[][]{dadina, ponti09, cappi2009, tombesi10}.

In this work, we present the first high-resolution X-ray spectrum of Mrk 509 (
M$_{BH}\sim1.02\times10^{8} M_{\odot}$, \citealt{li2024}) in the 2-10 keV band, obtained with the Resolve \citep{resolve} instrument onboard the X-Ray Imaging and Spectroscopy Mission \citep[\textit{XRISM},][]{xrism}. Leveraging the unprecedented spectral resolution of Resolve ($\approx$4.7\,eV at 6 keV, \citealt{resolve}), our analysis aims to disentangle the complex emission and absorption features in the Fe K band. This will allow us to place new, stringent constraints on the physical properties, dynamics, and location of the different gas flows, from the reverberating reflector to the transient inflows and outflows.

The approved \textit{XRISM} observation of Mrk 509 (P.I. M. Cappi) was designed with three primary scientific goals: \textit{i)} confirming and studying gas in- and out-flows, \textit{ii)} defining the origin of the neutral Fe K$\alpha$ line, and \textit{iii)} understanding the origin of the ionized iron complex. The \textit{XRISM} proposal triggered the organization of a broader, complementary multi-wavelength campaign involving \textit{NuSTAR}, \textit{XMM-Newton}, Swift, and optical telescopes (NTT/TNG) to provide a complete view of the source. However, in this work, we will focus mainly on the results obtained  analyzing the \textit{XRISM}/Resolve data and taking advantage of the \textit{XMM-Newton} \citep{jansen} and \textit{NuSTAR} \citep{harrison} capabilities to define the baseline X-ray continuum of Mrk 509. The broad-band properties of the source, including the spectral complexities in the soft energy band and the source variability, will be addressed in an upcoming paper (Missaglia et al., in prep.).


\section{Data reduction and analysis}

\begin{figure}
    \centering
    \includegraphics[width=1.10\linewidth]{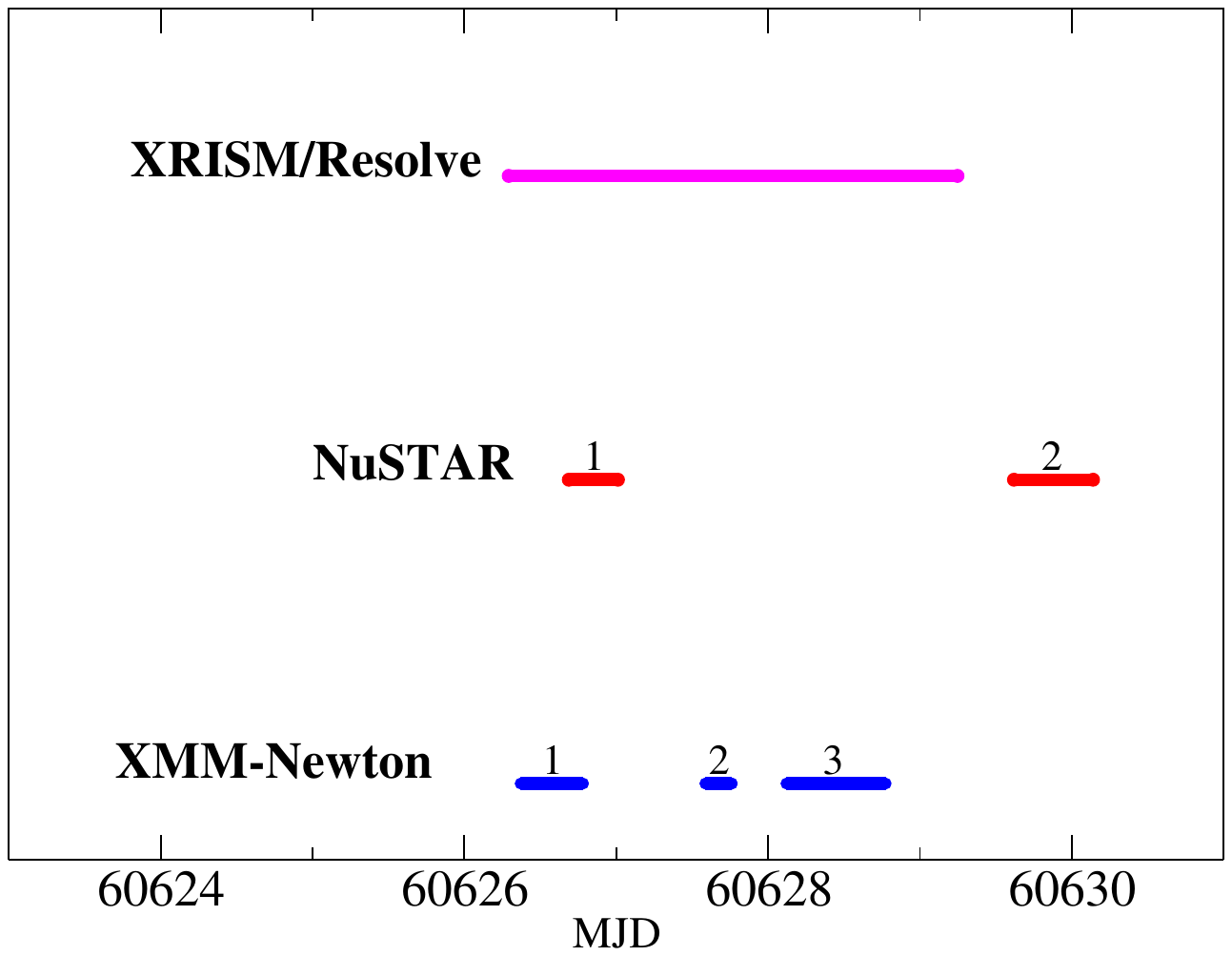}
    \caption{Time scheme of the X-ray observation presented in this work. The horizontal lines are representative of the total duration of the different observations.}
    \label{timelog}
\end{figure}

\textit{XRISM} observed Mrk 509 on November 11, 2024. The high-resolution Resolve calorimeter was operated in its standard PX NORMAL mode. We were unable to use the Xtend data for an anomaly in the camera during this observation. While prior studies with missions like \textit{XMM-Newton} and \textit{Chandra} have characterized Mrk 509 at CCD resolution, this work focuses primarily on the new, high-resolution Resolve spectrum. Nonetheless, we also present an analysis of the \textit{NuSTAR} and \textit{XMM-Newton} data collected via DDT observations triggered to coordinate with the \textit{XRISM}/Resolve pointing. These data are particularly important for properly constraining the continuum emission in the 2--10\,keV band. The source was observed twice with \textit{NuSTAR} in November 2024 for a total exposure of $\sim$34 ks. Quasi-simultaneous observations (three in total) were also performed with \textit{XMM-Newton}. Table~\ref{timelog} lists the details of all observations, while Fig. \ref{timelog} displays the temporal coverage of each pointing.

\begin{table}[]
    \centering

      \caption{Observations log with net exposure times}
         \label{obslog}
             \begin{tabular}{lc c}
            \hline
            \noalign{\smallskip}
            Instr\#     &  Obs. Starting Date& Exp (ks) \\
            &  \\
            \noalign{\smallskip}
            \hline
            \noalign{\smallskip}
            XRISM & 2024/11/12&106\\
            &\\
            NuSTAR\#1& 2024/11/12& 12.5\\
            &\\            
            XMM\#1& 2024/11/12& 17.2\\
            &\\ 
            XMM\#2 & 2024/11/13& 9.2\\
            & \\
            XMM\#3 & 2024/11/14& 30.7\\
            &\\
            NuSTAR\#2 & 2024/11/15& 21.2\\
            \noalign{\smallskip}
            \hline
         \end{tabular}
         \end{table}

\begin{figure}
   \centering
   \includegraphics[scale=0.35]{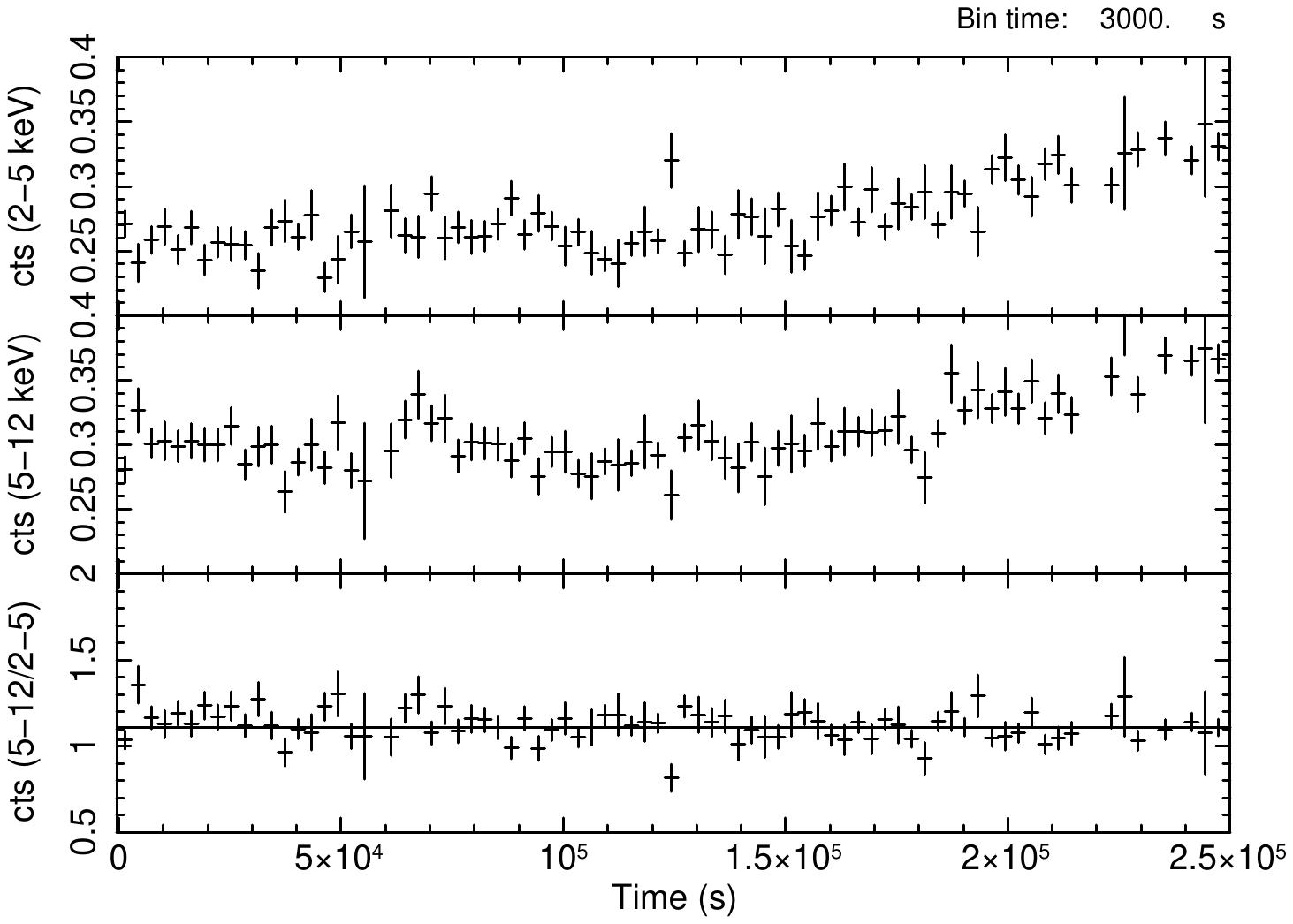}
      \caption{The non-background-subtracted \textit{XRISM}/Resolve light curves of Mrk 509 are shown in the 2–5 keV (top panel) and 5–12 keV (middle panel) energy bands. The bottom panel displays the corresponding hardness ratio. A clear brightening trend is evident toward the end of the observation, but it is not accompanied by a significant spectral variability.}
              \label{lc}%
    \end{figure}
The \textit{XRISM}/Resolve data were reduced following the \textit{XRISM} Quick-Start Guide v2.3, via HEASOFT version 6.34 and the associated calibration files. Briefly, the data were filtered to exclude anomalies in low-Earth orbit, and events from the calibration pixel (pixel 12) and pixel 27 with abnormal gain behaviour. The data were further screened to include only high-resolution primary (Hp) events. The technical report from the instrument team confirms the nominal functioning of the Resolve instrument during the observation, with an energy resolution of $\sim 4.6$\,eV at the Fe$^{55}$ reference energy, and no gain shift was detected over the course of the exposure. A light curve of the event list reveals flux variability, with a distinct brightening during the final third of the observation (see Fig.~\ref{lc}). A search for spectral variability, performed by analyzing the ratio between the 2–5 keV and 5–12 keV bands, suggests that no strong spectral variability is associated with this change in flux (see Fig.~\ref{lc}).  For this reason, a time-averaged spectral analysis is well-justified. The resulting spectrum is built with default bins of 0.5 eV, and the tools \textsf{rslmkrmf} and \textsf{xaarfgen} are used to make the redistribution matrix file (rmf) and ancillary response file (arf). We elected to create a ``Large'' (Gaussian core plus exponential tail \& Si K$\alpha$ instrumental line plus escape peak) rmf. After all screening (e.g., pixel-pixel coincident events) the net Resolve exposure time was approximately 106 ks. Finally, following the standard procedure (see for example \citealt{pds456}) the background was treated by directly incorporating the generated non-X-ray background (NXB) into our spectral models.

We processed \textit{XMM-Newton} EPIC-pn data using the Science Analysis System (SAS) v21.0. Calibrated event lists were generated from the raw observation data files (ODFs) using the standard pipeline chains. The camera was operated in Small Window mode to minimize photon pile-up. The resulting event lists were filtered to remove periods of high particle background flaring, and standard pattern selection criteria were applied.
For the spectral analysis, source counts were extracted from a 28$^{\arcsec}$-radius circular region centered on the source's position. The background spectrum was estimated from two nearby, source-free 26$^{\prime\prime}$-radius circular regions located on the same CCD. The corresponding arfs and rmfs were generated for spectral fitting.

Similarly, we processed \textit{NuSTAR} data for both observations using the NuSTAR Data Analysis Software (NuSTARDAS) package. Source spectra were extracted from a 1.5$^{\prime}$-radius circular region centered on the target. For the background, we used a separate 1.88$^{\prime}$-radius circular region located on the same detector, chosen to be free of source contamination. The \textsf{nuproducts} task was used to filter the data, apply standard screening criteria, and generate all final data products, including spectra and response files.

\begin{figure}
   \centering
\includegraphics[scale=0.37]{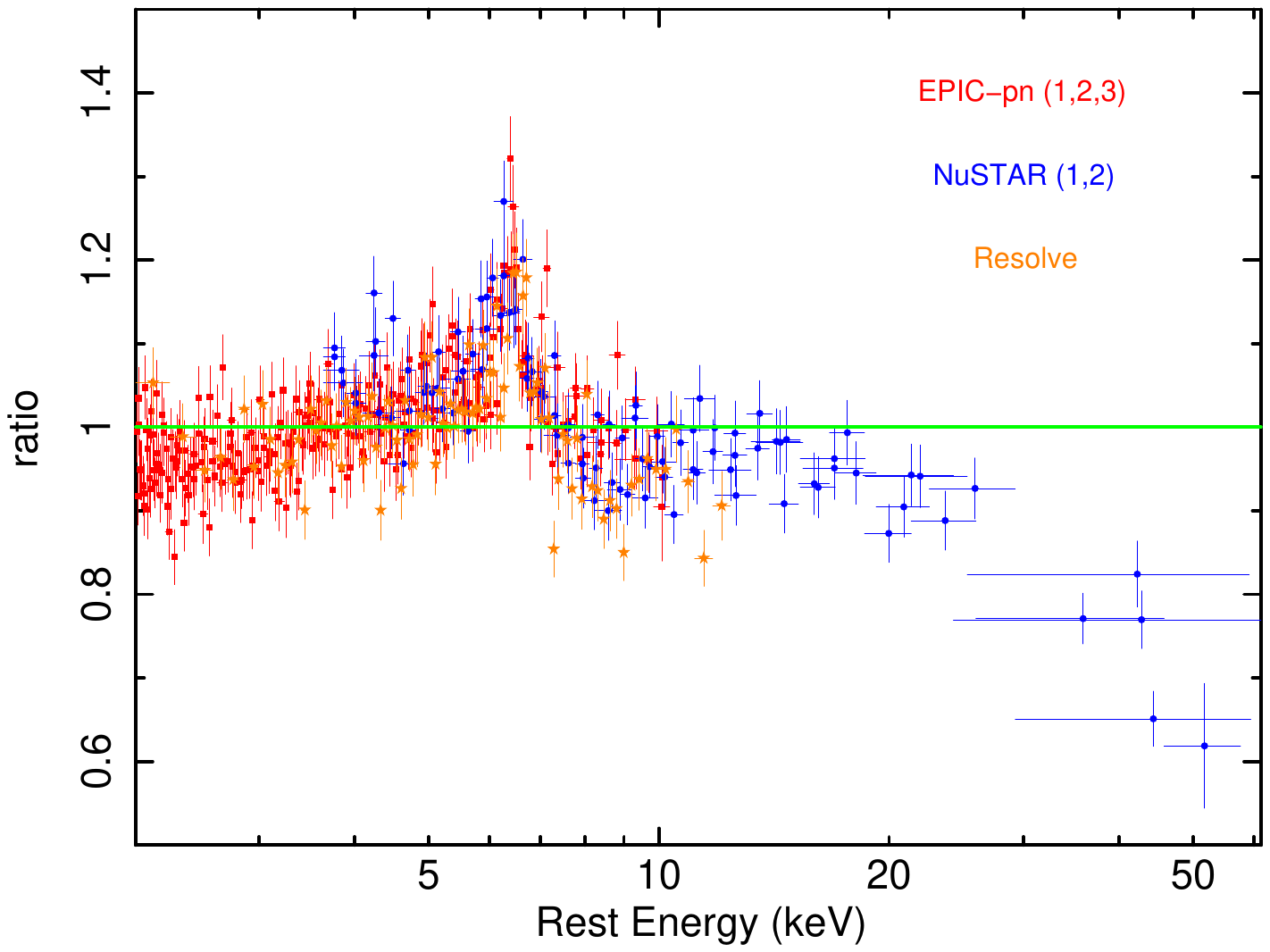}
\includegraphics[scale=0.37]{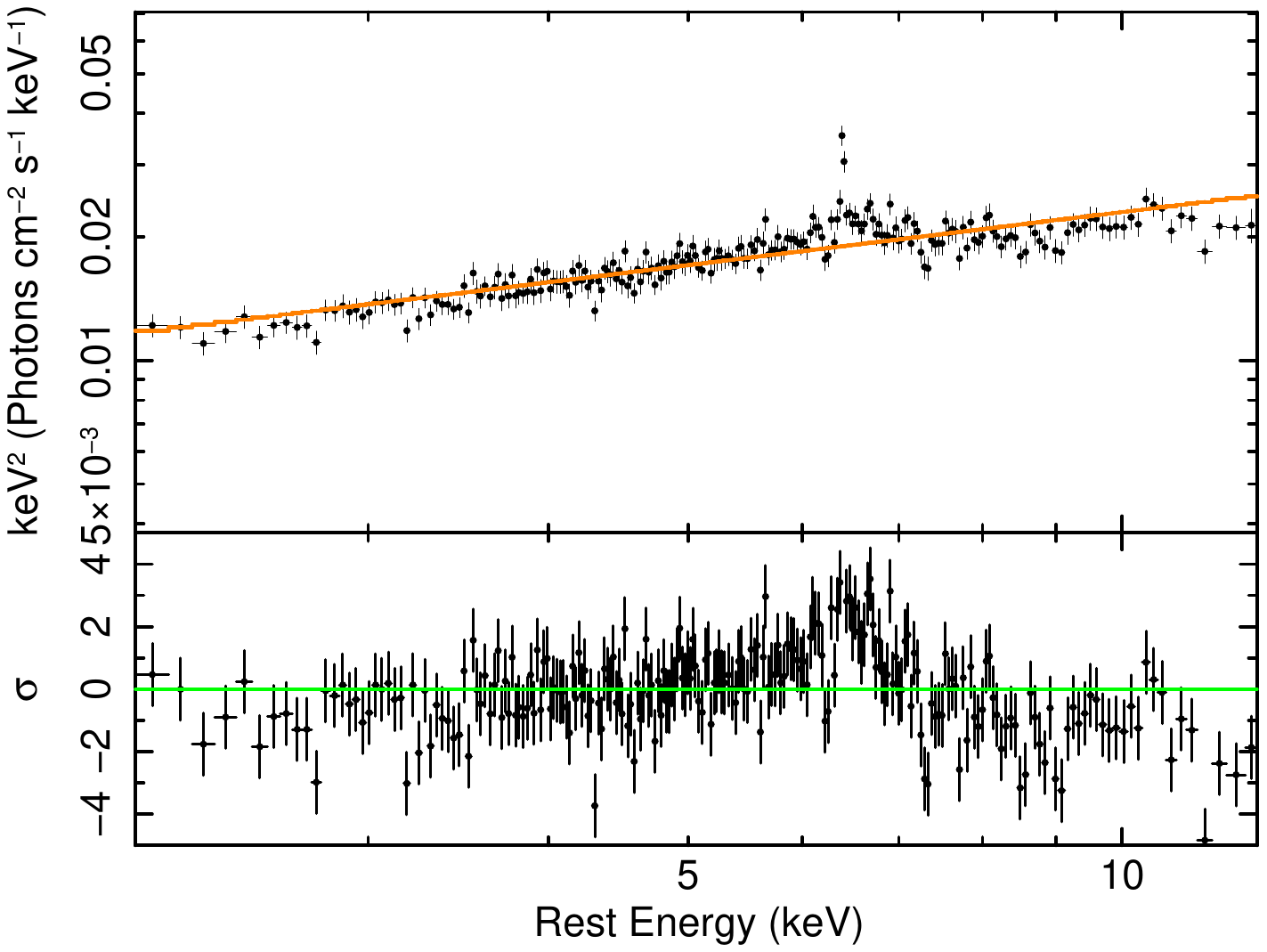}
   \caption{ {\it Upper panel}: the 2-60 keV ratio between the data and a simple power-law model for the Mrk 509 X-ray data taken in this campaign (see Tab. \ref{obslog}), expressed in terms of standard deviations. Here, for clarity purposes, data have been strongly rebinned (each data point has a 25$\sigma$ significance).
   {\it Lower panel}: 2–12 keV spectrum of Mrk 509 observed by \textit{XRISM}/Resolve. The orange line is the best fit using a simple power-law model.
   Also in this case, Resolve data have been rebinned (each data point here has a significance of 15$\sigma$) for visual clarity. }
              \label{spe0}%
    \end{figure}

To prepare the data for spectral fitting, we applied specific grouping and bandpass selections for each instrument. The \textit{XRISM}/Resolve data were analyzed in the 2--12\,keV band to avoid calibration uncertainties at the spectral boundaries. Following \citet{mizumoto26}, these data were uniformly grouped by a factor of 10, resulting in a final bin size of 5\,eV \citep{xrism2025}. Although this slightly overbins the nominal energy resolution of Resolve, we adopted this conservative approach to enhance the signal-to-noise ratio (S/N) per spectral bin, facilitating both the detection and the modeling of weak and broad spectral features. Furthermore, the choice of a constant energy binning was dictated by the mathematical requirements of certain convolution models used in our analysis, such as \texttt{gsmooth} (see Sect. \ref{xrism}). Finally, we anticipate here that this choice has not strong impact on our data analysis, since the width of the narrow Fe emission line is found to be of the order of $\sigma\sim$10\,eV (see Sect. \ref{xrism}). The EPIC-pn data were grouped to a minimum of 100 counts per bin, while the \textit{NuSTAR} data were similarly grouped to ensure a minimum of 50 counts per bin.

\section{Defining a baseline model using the 2-60 keV XMM-Newton and NuSTAR data}
\label{broadband}

A simple power-law model to simultaneously fit the 2-60 keV \textit{XRISM}/Resolve, the \textit{XMM-Newton}/EPIC-pn and \textit{NuSTAR} data reveals several key features (see Fig.~\ref{spe0}, upper panel). The most prominent are: a) a narrow emission line centered at E $\approx$ 6.4 keV, and b) a broad excess spanning the 4–7 keV band. 
\begin{table*}
  \centering  
      
     \caption{ Best-fits of the \textit{XMM-Newton} EPIC-pn plus \textit{NuSTAR} spectra only: 
    \textit{Upper Table} - Col. I: Model name; Col. II: Photon index of the primary emission; Col. III: Column density of the partial absorber; Col. IV: Covering factor; Col. V: Energy of the emission line; Col. VI: direct-to-reflected continuum ratio; Col. VII: Width of the emission line; Col. VIII: Equivalent width of the emission line; Col. IX: Cash statistics/degrees of freedom. \textit{Lower Table} - Col. I: Model name; Col. II: Column density of the partial absorber; Col. III: Covering factor; Col. IV: Photon index of the primary emission; Col. V: Inner radius of the accretion disk; Col. VI: Ionization of the accretion disk; Col. VII: Reflection fraction parameter; Col. VIII: Spin of the BH; Col. IX: Cash statistics/degrees of freedom. Cols. X-XIII: cross-normalization constants between \textit{XMM-Newton} EPIC-pn and \textit{NuSTAR}. Reported errors are at 90\% confidence level.}  
 \tiny
 \addtolength{\tabcolsep}{-0.35em}
        \begin{tabular}{l c c c c c c c c c c c c c }

\hline\hline
&&&&&&&&&\\
Model & $\Gamma$ &N$_H$& C$_f$& E$_{FeK\alpha}$ & R & $\sigma$ & EW & C/dof & \sffamily{C$\textsubscript{pn,2}$} & \sffamily{C$\textsubscript{pn,3}$} & \sffamily{C$\textsubscript{FPMA/B,1}$} & \sffamily{C$\textsubscript{FPMA/B,2}$} \\
&&&&&&&&&\\
&&(10$^{22}$ cm$^{-2}$)& & (keV) &  &(eV) & (eV) &\\
&&&&&&&&&\\
\hline
&&&&&&&&&\\
PC1& 1.65$\pm0.02$ & 12.5$\pm$1.7 & 0.23$\pm0.02$& 6.39$\pm0.03$ & - & 102$\pm$40 & 68$\pm$3  & 2555/2422 & 1.03$\pm$0.01 & 1.11$\pm$0.02 & 1.24$\pm$0.02 & 1.27$\pm$0.02 \\        
&&&&&&&&\\
PC2 & 1.76$\pm0.02$ & 12.1$\pm1.2$ & 0.31$\pm0.03$& - & 0.38$\pm0.04$  & - & - & 2529/2424 & 1.03$\pm$0.01 & 1.11$\pm$0.01 & 1.24$\pm$0.02 &1.27$\pm$0.02 \\ 
&&&&&&&&\\

\end{tabular}

\begin{tabular}{l c c c c c c c c c c c c }
\hline\hline
&&&&&&&\\
Model & N$_{H}$ &  C$_{f}$ &  
$\Gamma$& r$_{in}$ & log $\xi$ & R & a & C/dof & \sffamily{C$\textsubscript{pn,2}$} & \sffamily{C$\textsubscript{pn,3}$} & \sffamily{C$\textsubscript{FPMA/B,1}$} & \sffamily{C$\textsubscript{FPMA/B,2}$} \\
&&&&&\\
&(10$^{22}$ cm$^{-2}$)& &   &(ISCO) & &&\\
&&&&&\\
\hline
&&&&&\\
RR1  & -  & - &   1.47$^{+0.01}_{-0.01}$ & 1* & 4.28$^{+0.05}_{-0.09}$ & 0.74$^{+0.48}_{-0.21}$ & $>$0.85 & 2683/2422 & 1.03$\pm$0.01 & 1.11$\pm$0.01 & 1.24$\pm$0.02 &1.27$\pm$0.02 \\ 
&&&&&&\\
RR2& 10.35$\pm$2.10 & 0.20$^{+0.04}_{-0.02}$ & 1.59$^{+0.08}_{-0.04}$  & 1* & 3.56$^{+0.20}_{-0.68}$ & 0.41$^{+0.39}_{-0.24}$ & 0.99$^u$ & 2519/2420 & 1.03$\pm$0.01 & 1.11$\pm$0.01 & 1.24$\pm$0.02 &1.27$\pm$0.02  \\
&&&&&&\\
\hline
\end{tabular}
 \tablefoot{Models RR1-RR2 have zgauss with energy and $\sigma$ fixed (6.4 keV and 20 eV).\\ Model PC1: constant$\times$TBabs$\times$zpcfabs(cutoffpl + line);\\ Model PC2: TBabs(cutoffpl+pexmon+(1+constant)$\times$zTBabs(cutoffpl));\\ Model RR1: constant$\times$TBabs(zgauss+relxill) with Rbr=15, AFe=1;\\ Model RR2: constant×TBabs×zpcfabs(relxill+zgauss).
 }
  \label{BB} 
\end{table*}

To determine the best modeling of the continuum, we started by analyzing the 2–60 keV data from \textit{XMM-Newton}/EPIC-pn and \textit{NuSTAR} only. At this stage, we excluded the data below 2 keV to avoid the complexities associated with the known strong soft excess \citep[see e.g.,][]{mehdip2011,petrucci2013,petrucci} and the complex warm absorber \citep[][]{kaastra} that strongly affect the low energy (E $\leq$ 2 keV) spectrum of Mrk 509.

The spectral analysis, including all \textit{NuSTAR} and EPIC-pn datasets collected during our campaign (total of 5 observations, see Table~\ref{obslog}), was performed using the \textsc{Xspec} fitting package v12.15.1\citep{xspec}. We included a cross-normalization constant to account for differences in absolute calibration between the instruments, as well as for intrinsic source variability (see columns X-XIII in Table \ref{BB}). Mrk 509 is known to exhibit X-ray flux variations on typical timescales of days \citep{kumari}. This behavior is indeed confirmed by the \textit{XRISM}/Resolve light curves (see Fig.~\ref{lc}).

We started using a simple cut-off power-law model plus a Gaussian emission line to reproduce the Fe~K$\alpha$ feature. The high-energy cut-off was fixed at a fiducial value of $E_c=150\,\rm keV$ \citep[see][]{petrucci2013}, as the current data do not provide meaningful constraints on this parameter. We fixed the line width to $\sigma=20$\,eV, consistent with the narrow feature clearly resolved in the \textit{XRISM}/Resolve data (see both panels of Fig.~\ref{spe0}).

The energy of the Gaussian is found to be consistent with neutral Iron (E=6.39$\pm$0.02). If we try to model the remaining curvature observed in the 4-7 keV band with a broad Gaussian, we found that its centroid is at E$\sim$5.4 keV  and it has a width of $\sigma\sim$1.5 keV, thus indicating that most probably the underlying continuum is poorly described by the simple power-law modeling. We have also tried to fix the broad Gaussian to be at the same energy of the narrow component, but in this case residuals are still present.

To describe the residual spectral curvature observed at 4-7 keV, we thus explored different scenarios involving a different modeling of the continuum: i) the presence of a neutral, partial absorber (models PC1 and PC2 in Table~\ref{BB}) ; and/or ii) a relativistic scenario due to reflection from material in the disk (models RR1-RR2 with \texttt{relxill} in Table~\ref{BB}).

\begin{figure}
   \centering
\includegraphics[scale=0.37]{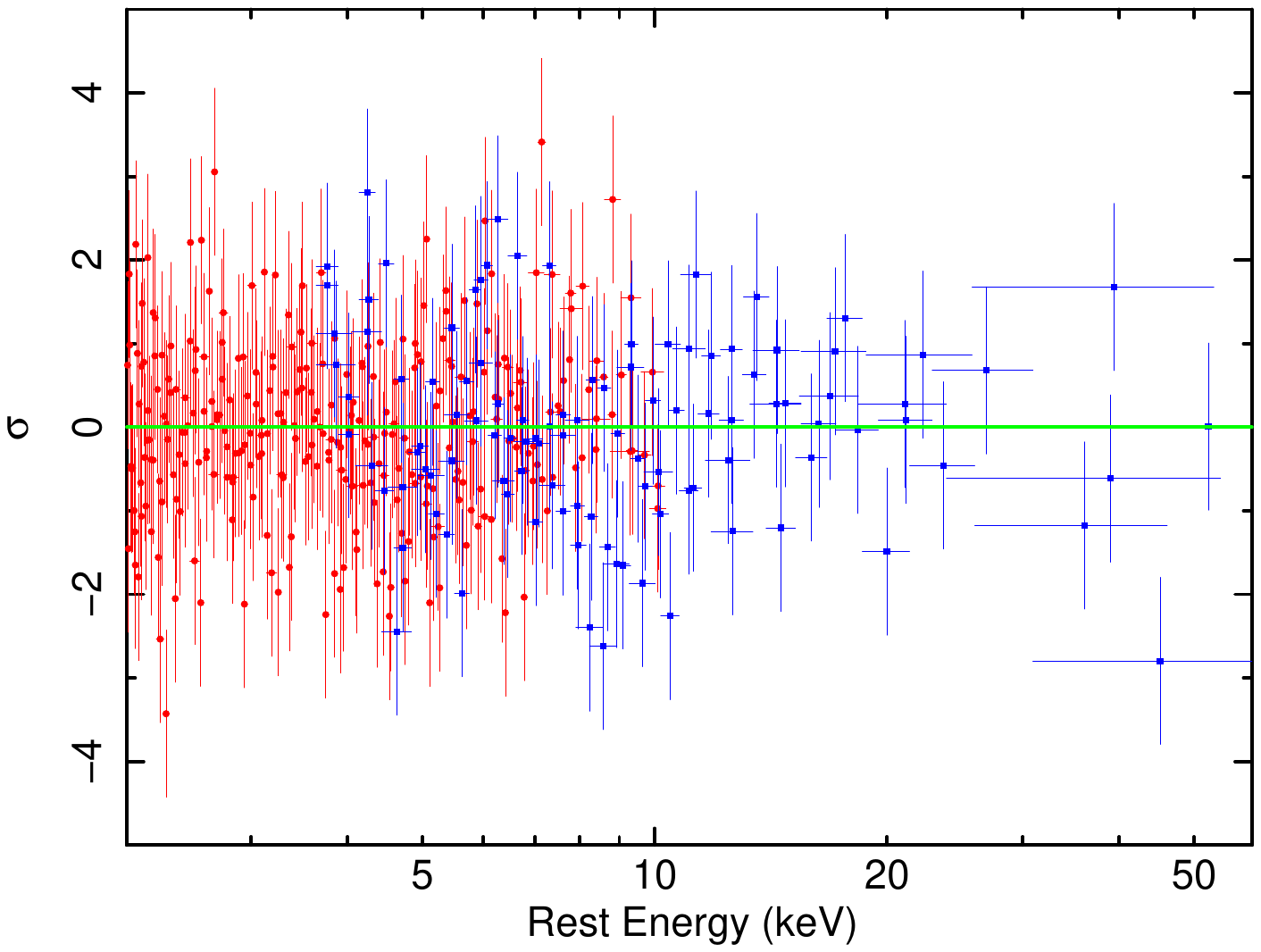}
\includegraphics[scale=0.37]{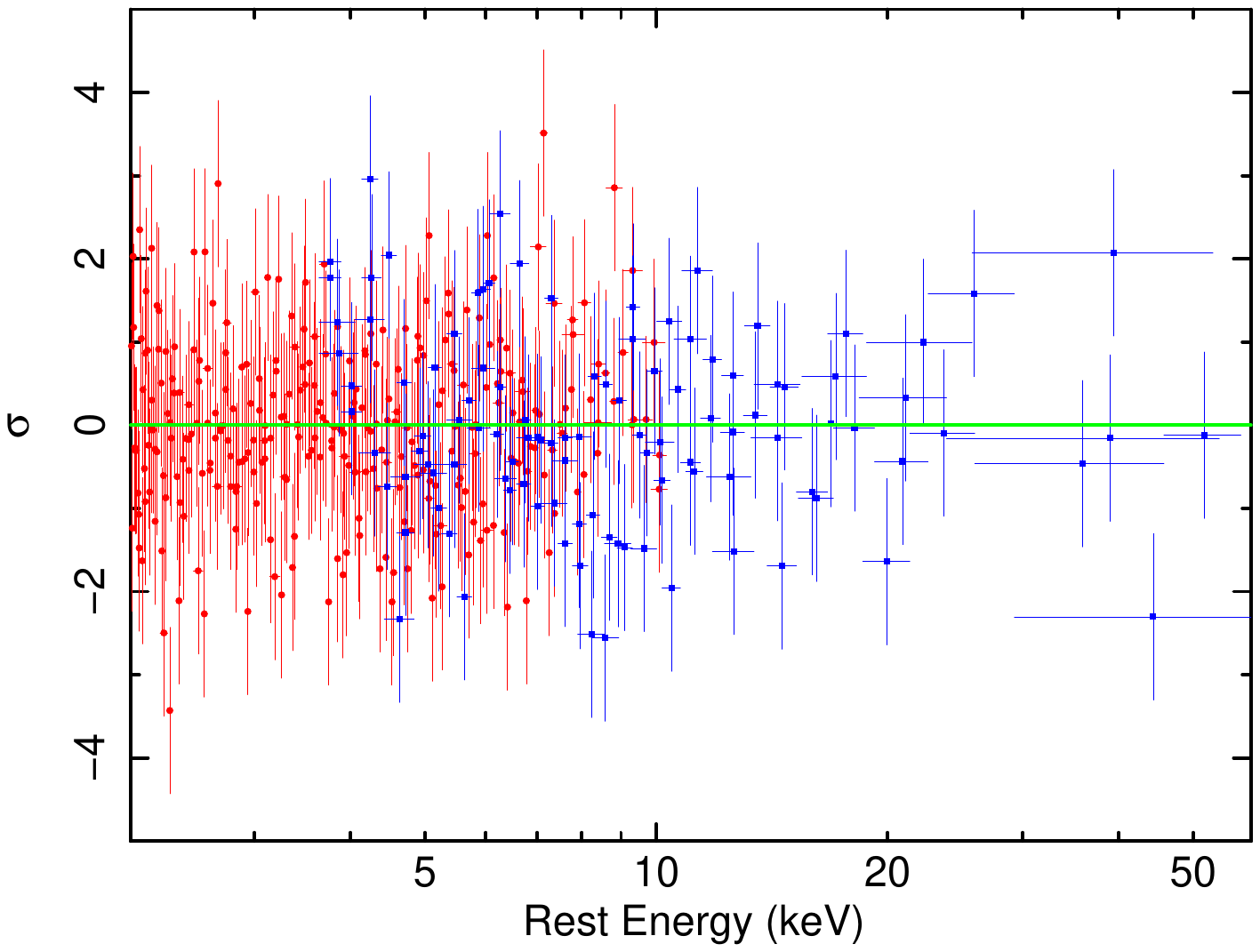}

   \caption{\textit{Upper panel}: 
   Residuals of model PC2 (neutral partial covering with distant reflection component). \textit{Lower panel}: residuals of model RR2 (neutral partial covering with relativistic reflection component and narrow Fe K$\alpha$ Gaussian emission line). 
   In all panels \textit{XMM-Newton} data are in red, while \textit{NuSTAR} data are in blue. }
              \label{pexpart}%
    \end{figure}

\begin{figure}
   \centering
\includegraphics[scale=0.35]{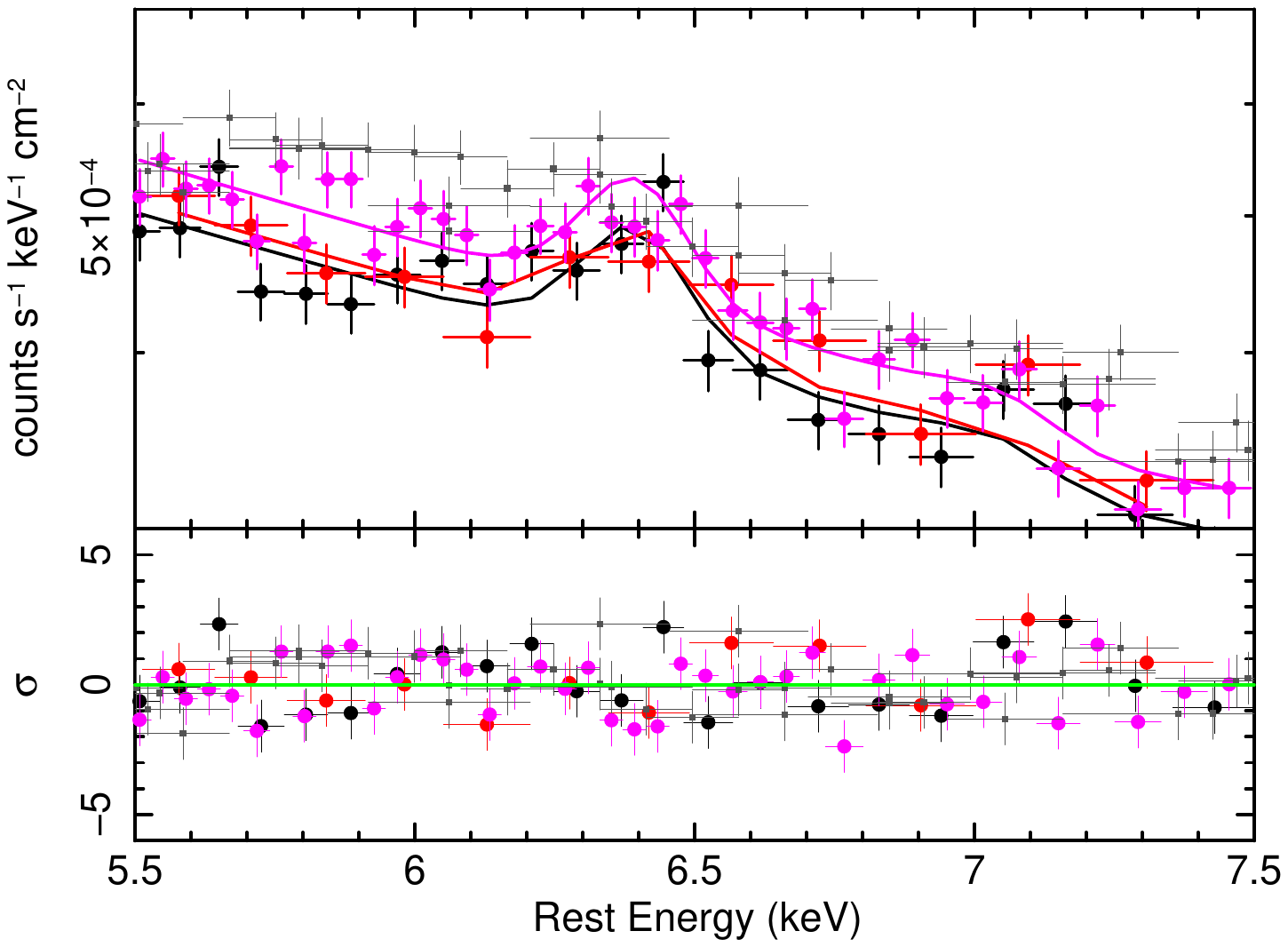}
\caption{Zoom in the 5.5-7.5 keV range model PC2 (see Table~\ref{BB}). XMM-Newton data observation 1 are in black, observation 2 in red, observation 3 in magenta. NuSTAR data are all in grey. There are no significant positive/negative residuals.}
              \label{zoom}%
    \end{figure}

The partial covering models provide a robust description of the broadband data. Specifically, when a simple Gaussian line is added, the feature is found to be consistent with emission from cold material (see model PC1 in Table~\ref{BB}). Furthermore, motivated by the \textit{XRISM}/Resolve spectra (see Sect.~4)—which clearly reveal a narrow core—we tested a scenario where this component originates from neutral reflection from distant material (modeled using \texttt{pexmon}, \citealt{nandra07}). This approach significantly improves the quality of the fit (see model PC2 in Table~\ref{BB}). Finally, we tested the possible presence of an additional broad Fe K$\alpha$ Gaussian component, finding that it is not statistically required when the partial covering model is adopted. This modeling, moreover, allows us to recover photon index values ($\Gamma\sim1.7$) in agreement with those previously reported in literature ($\Gamma\sim1.6$--$1.7$; \citealt{pounds, ponti09, ponti}) and with the average value found in AGN samples \citep{nandra94, piconcelli05, dadina07, dadina08, bianchi09, ricci17, middei19}.

To test the relativistic reflection scenario, we initially applied a model consisting of a relativistic reflection component (\texttt{relxill}, \citealt{relxill}) and a narrow Fe~K$\alpha$ Gaussian emission line. As done in the previous case, we fixed the high-energy cut-off to be E$_c=150$\,keV and, as shown in \cite{petrucci2013}, we fixed the inclination angle of the system to be $\Theta=30^{\circ}$. This baseline model provides a poor description of the data (model RR1 in Table~\ref{BB}), leaving significant positive residuals. Similarly to the simple power-law modeling, the inclusion of a partial covering component is required as well (model RR2 in Table~\ref{BB}). This addition significantly improves the fit (see residuals in the lower panel of Fig.~\ref{pexpart}). Left free to vary, the black hole spin (indicated by the a parameter in Table \ref{BB}) remains unconstrained, while the reflection component is consistent with that measured in the PC2 model. Following the approach adopted for model PC2, we also tested a scenario where the narrow Fe~K$\alpha$ line originates from the distant torus (modeled using \texttt{pexmon}). This resulted in a significantly worse description of the data compared to model RR2. Finally, it is worth noting that in all these relativistic scenarios, the recovered photon index of the primary emission is consistently flat (see Table~\ref{BB}).

As reported in Table~\ref{BB}, the PC2 and RR2 models provide the best fits to the broadband data (see also both panels of Fig.~\ref{zoom} for a zoom-in of the residuals in the Fe K band for the PC2 and RR2 best fits). The RR2 scenario introduces four additional free parameters but provides only a marginal statistical improvement ($\Delta$C=10) and, crucially, the PC2 model drives the primary photon index to values perfectly in agreement with those typically observed in Seyfert galaxies \citep{nandra94, ricci17}. In the latter scenario, we assume that the reflection component originates in the torus, as suggested by the narrowness of the Fe K features (see also Section~\ref{xrism}). Consequently, in our modeling, this component is not affected by the partial absorber, which is assumed to cover only the emission originating in the Mrk~509 core.

\begin{figure}
   \centering
\includegraphics[scale=0.33]{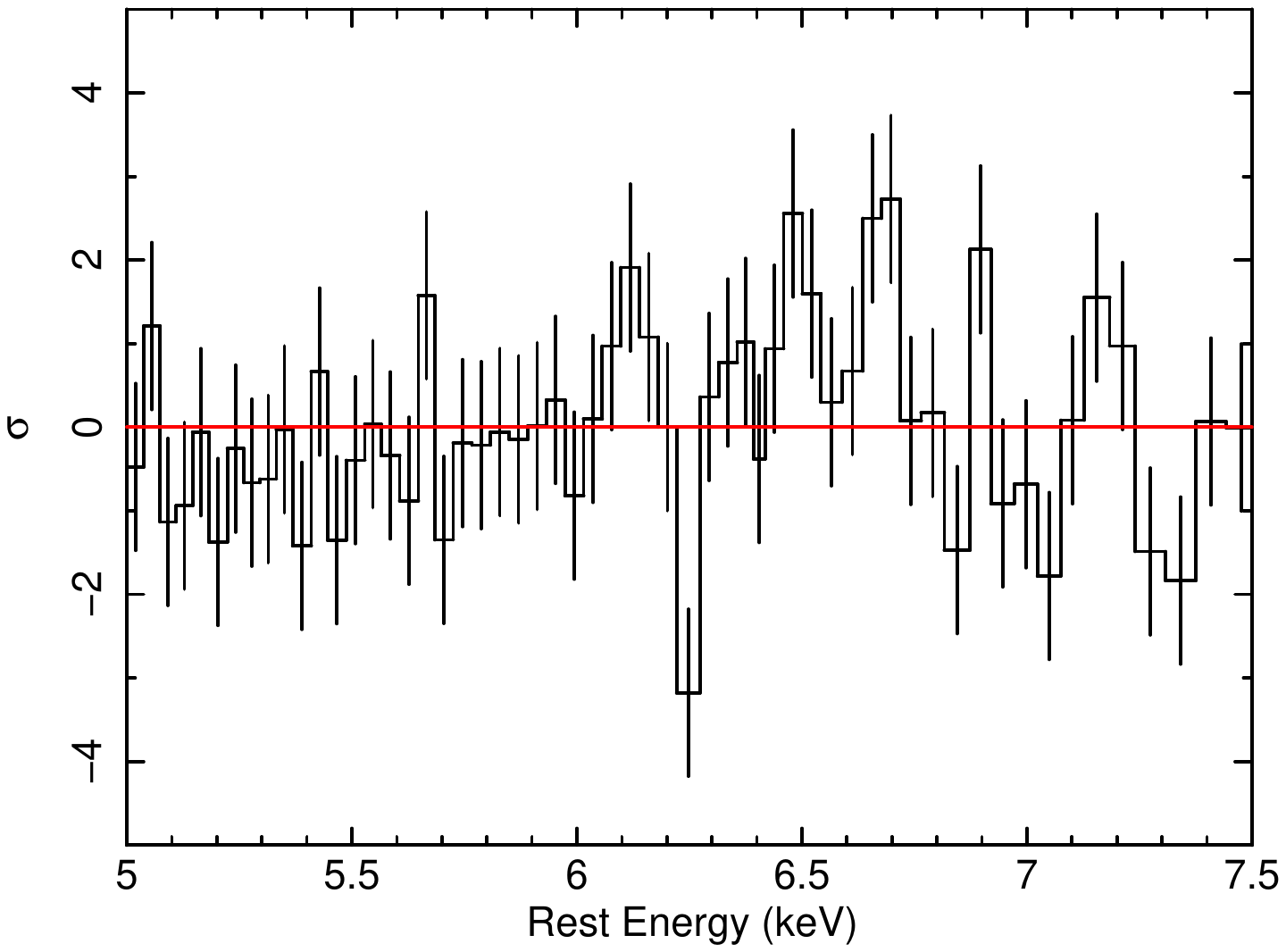}

   \caption{5--7.5 keV zoom in of the Resolve residuals of the model including a cold partial covering and the zFeKlor component to model the narrow Fe K$\alpha$ complex. Data have been strongly rebinned for clarity purposes (each data point in the plot has a 20$\sigma$ significance) . The strongest departures from the model are the possible absorption feature at E$\sim$6.2 keV and a broader component in 6-6.8 keV range.  
   }
              \label{model0}%
    \end{figure}

\begin{figure}
    \centering
   \includegraphics[scale=0.35]{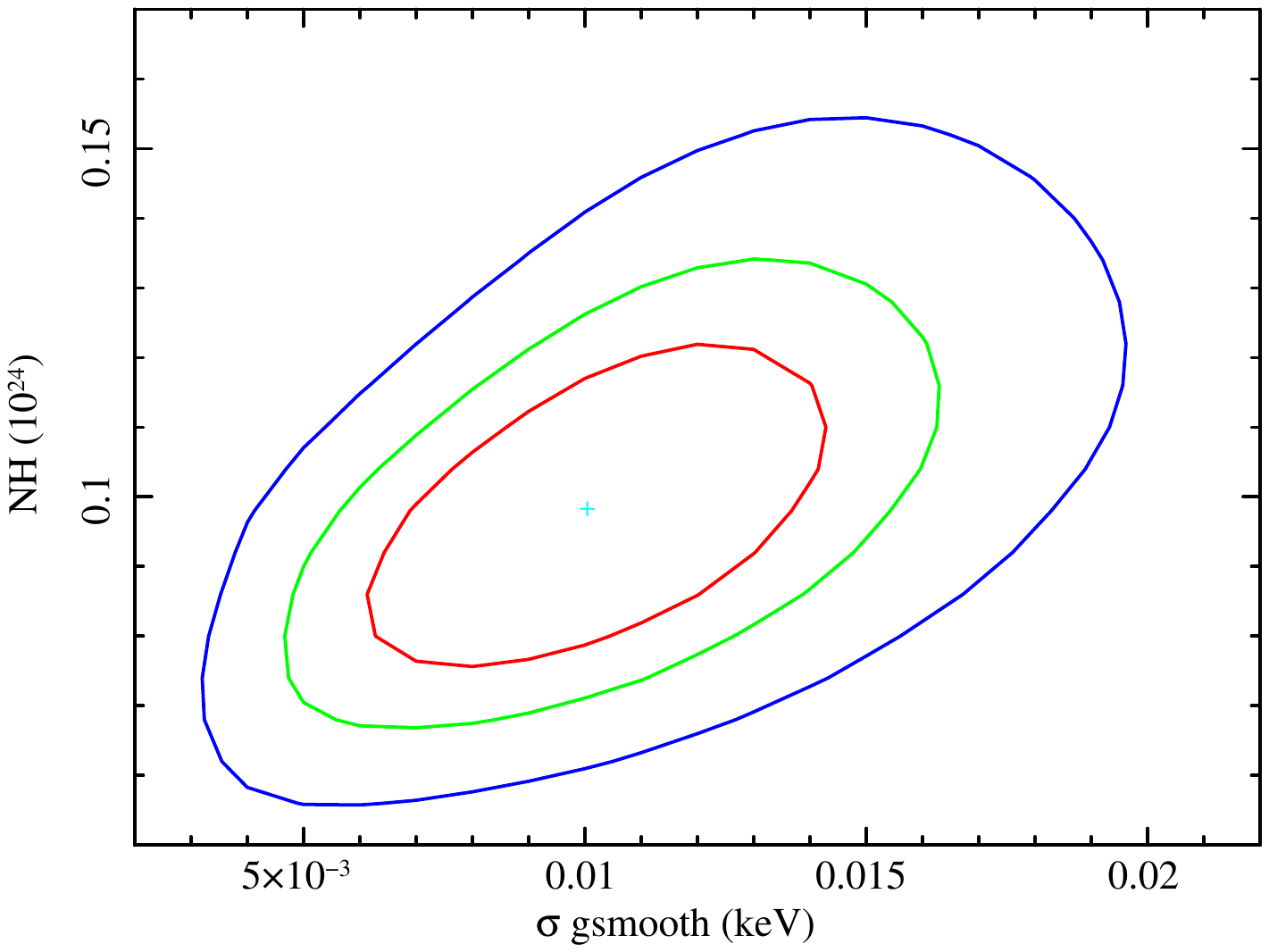}
   \caption{Confidence contours between the column density and broadening kernel for the distant reflector as measured using MYTorus modeling of the reflector and associated lines (see Section~\ref{refl})}
                 \label{torus}
\end{figure}

\begin{figure}
   \centering
\includegraphics[scale=0.35]{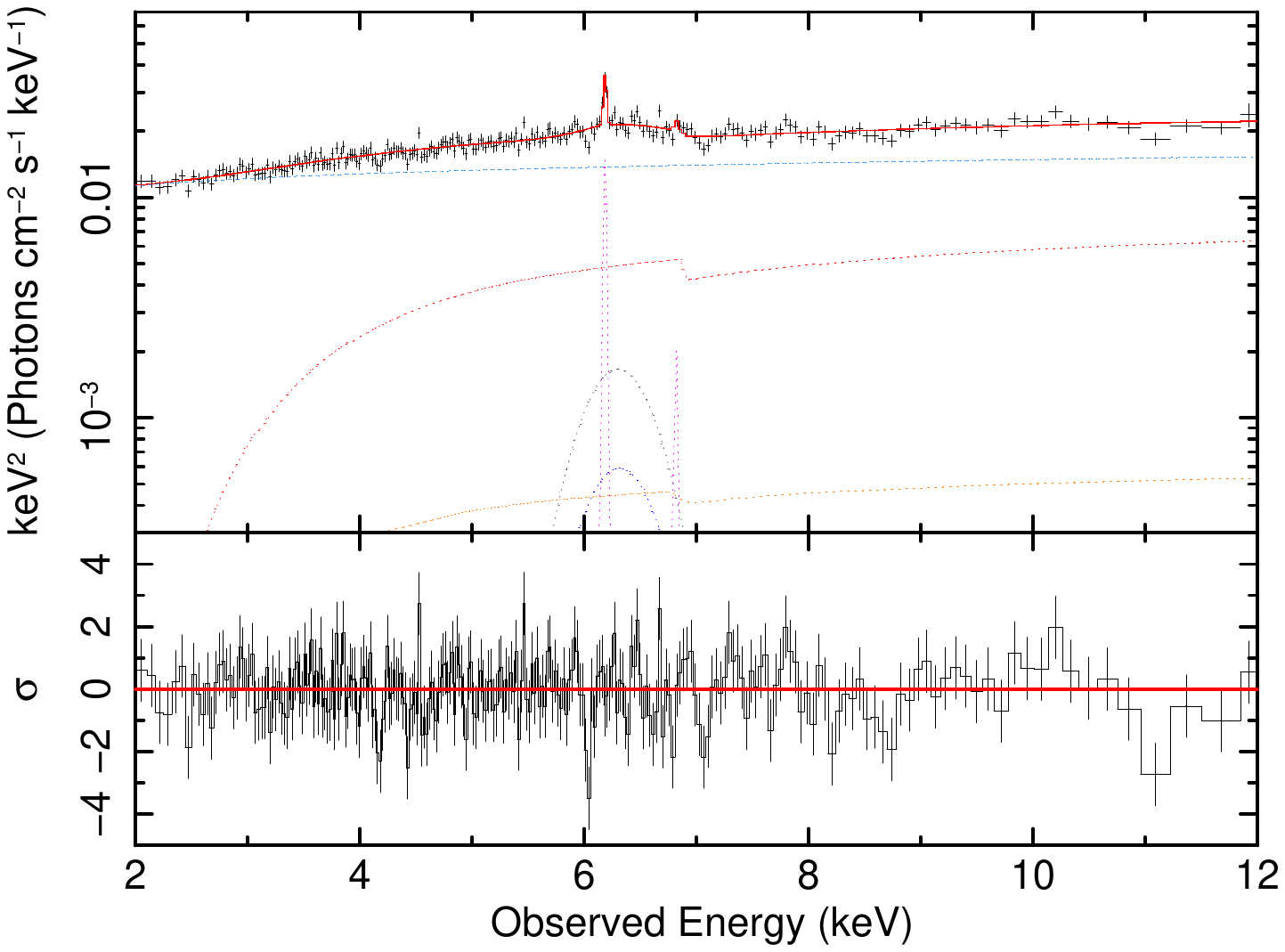}

   \caption{2–12 keV spectrum of Mrk 509 modeled with a partial covering of the primary emission and broad Gaussian line plus neutral reflection (\texttt{MYTorus} model, see model I in Tab.~\ref{pcmy}). Components are in dotted lines: absorbed powerlaw is in red, absorbed broad Gaussian line in blue, MYTorusL in magenta, MYTorusS in orange. 
   }
              \label{zpc}%
    \end{figure}

\section{The XRISM/Resolve view of the Fe complex}
\label{xrism}

Building on the results presented above, we tested both the simple partial covering model (named PC in the following) and the relativistic scenario (named RR in the following) on our \textit{XRISM}/Resolve data. Starting with the PC2 scenario, we used a simple \texttt{zpcfabs} model (see Section~\ref{broadband}), yielding a canonical photon index of $\Gamma\sim1.8$, in excellent agreement with the values obtained from \textit{XMM-Newton} EPIC-pn and \textit{NuSTAR} data. In this scenario, the primary emission is partially absorbed (covering fraction $C_f \sim 0.3$) by a neutral component with a column density of $N_H \sim 1.3\times10^{23}$\,cm$^{-2}$ (again consistent with the broadband fit obtained in Section~\ref{broadband}).  

The narrow Fe K$\alpha$ line has been modeled with the \texttt{zFeklor} model \citep[this model includes seven Lorentzian profiles to describe the line, see][]{holzer97}. The energy centroid of this model is not a free parameter of the fit and when we left its redshift free to vary the best fit is obtained at $z=0.03389^{+0.00072}_{-0.00025}$ (errors correspond to the 90\% confidence level here and throughout the paper, unless stated otherwise), well in agreement with the redshift of the source. To estimate the line width, we applied a \texttt{gsmooth} convolution model to the \texttt{zFeklor} component, yielding $\sigma_n=10\pm5$\,eV and equivalent width $EW_n\sim27$\,eV. These parameters strongly suggest that this feature originates far from the central SMBH. 

Some residuals are still present at the rest energy of the Fe K$\alpha$ complex, between 6.5 and 6.7 keV (see Fig.~\ref{model0}). Following previous results on Mrk~509 \citep{ponti09,ponti,yaqoob2003,yaqoob2004} and the features observed in NGC 4151 \citep{XRISM4151}. , we added a broad Gaussian component that improves the fit ($\Delta C\sim20$ for three parameters of interest). The line parameters are found to be: $E_b\simeq 6.49\pm0.15$\,keV, width of $\sigma_b\simeq320\pm220$\,eV and $EW_b\simeq 90\pm50$\,eV. The width of this feature (v$\sim$14800 km/s) suggests an origin close to the SMBH, corresponding to r$\sim50\,r_g$.

Interestingly, very similar results are obtained in the RR scenario (see model RR-I in the lower panel of Table~\ref{pcmy}). Assuming a maximally spinning SMBH and a standard radial emissivity profile ($\epsilon \propto  r^{-3}$), we derive a lower limit for the inner edge of the reflection disk of r$_{in}\geq27\,r_g$ (consistent with what obtained with the broad Gaussian), alongside a low reflection fraction of R$\sim$0.17 and a consistent broadening of the narrow FeK line.

\subsection{Characteristics of the distant reflection}
\label{refl}

As seen in Tab. \ref{pcmy}, the properties of the narrow emission line are consistent with an origin far from the SMBH, using both the PC and the RR scenarios. In order to achieve a more robust physical characterization, in the PC scenario we used the \texttt{MYTorus} model \citep{mytorus, yaqoob12} to account for this component.  \texttt{MYTorus} was infact designed specifically for modeling the X-ray spectra of AGN and their associated fluorescent lines when originate in the dusty torus. Specifically, we added the \texttt{MYTorusL} (Fe K$\alpha$ and Fe K$\beta$ fluorescent emission lines, latest available version) and \texttt{MYTorusS} (reflected continuum) components to the primary power-law continuum, linking their column densities to ensure self-consistency and fixing the inclination angle at $\Theta=0^{\circ}$, therefore assuming matter out of the line of sight.  Not only \texttt{MYTorusL} reproduces the narrow Fe line profile (similar to \texttt{zFeklor}), but it crucially ensures the proper treatment of the Compton shoulder. The photon index of the \texttt{MYTorus} components was tied to that of the primary power-law to maintain a uniform spectral slope. Since a preliminary Gaussian fit showed the narrow Fe line energy to be consistent with 6.4\,keV, we fixed the redshift of the \texttt{MYTorus} components to that of Mrk~509, assuming no additional bulk motion. Finally, to test for line broadening, we applied a Gaussian smoothing kernel (\texttt{gsmooth} in \textsc{Xspec}) to the \texttt{MYTorusL} component.

As displayed in Fig.~\ref{torus}, the material responsible for the narrow emission lines (Fe~K$\alpha$ and Fe~K$\beta$) must be located far from the SMBH, given that the line broadening is of the order of $\sigma=10\pm5$\,eV. This corresponds to a velocity of $v\sim1000\pm500$\,km\,s$^{-1}$ (FWHM), a result that is well in agreement with what obtained using zFeklor models and implying a distance on pc scales for a SMBH with a mass of $M\sim10^8\,M_{\odot}$ \citep{peterson04,li2024}. This result strongly supports the hypothesis that the Fe~K lines originate in the torus \citep{antonucci1993}.

On the other hand, the column density of this component is measured to be relatively low ($N_H\sim10^{23}$\,cm$^{-2}$) compared to the average value ($N_H\sim10^{24}$\,cm$^{-2}$) found in samples of Compton-thin AGN using \textit{NuSTAR}, but in good agreement with the lower end of the column densities measured for the Compton-thin sources in the sample \citep{zhao2021}. 
It is worth noting here that adding \texttt{MYTorus} in the RR scenario does not meaningfully improve the fit.

\begin{table*}[h]
\scriptsize
    \caption{Partial covering models}
    \label{pcmy}
    \begin{subtable}{\linewidth}
        \caption{\textit{PC models}: Col. I: Model name; Col. II: Photon index; Col III: Column density of the distant reflector; Col IV: Covering factor; Col. V: Width of the MYTorus line (gsmooth $\sigma$); Col VI: Energy of the broad Gaussian line in emission; Col. VII: Width of the broad Gaussian line; Col. VIII: Equivalent width of the broad Gaussian line; Col. IX: Column density of the partial covering component; Col. X: Ionization factor; Col XI: Observed redshift of the ionized absorber; Col. XII: Cash statistic value/degrees of freedom; Col. XIII: Akaike Information Criterion.}
        \begin{tabular}{lccccccccccccc}
\hline
\noalign{\smallskip}
\hline
\noalign{\smallskip}
Model & $\Gamma$ & $N_{H,torus}$ & $C_f$ & $\sigma_n$ & $E_{Fe,b}$ & $\sigma_b$ & EW$_b$& $N_{H,pc}$ & $\log\xi$ & $z_{obs}$ & C/dof & $\Delta$AIC \\
\noalign{\smallskip}
 & & ($10^{23}$\,cm$^{-2}$) & & (eV) & (keV) & (eV) & (eV) & ($10^{23}$\,cm$^{-2}$) & & & &\\
\hline\noalign{\smallskip}
I & $1.84^{+0.08}_{-0.08}$ & $1.00^{+0.12}_{-0.07}$ & $0.32^{+0.19}_{-0.17}$ & $10.2^{+4.5}_{-4.3}$ & $6.489^{+0.116}_{-0.148}$ & $324^{+224}_{-179}$ & 87$^{+52}_{-49}$& $1.44^{+0.47}_{-0.52}$   & & & 1917.2/1988 & \\
&&&&&&&&&&&&\\
n8 & $1.80^{+0.07}_{-0.06}$ &$0.98^{+0.12}_{-0.07}$  & $0.29^{+0.15}_{-0.13}$ & $10.1^{+4.6}_{-4.4}$ & $6.502^{+0.193}_{-0.131}$ & $456^{+150}_{-127}$ & 125$_{-51}^{+55}$ & $3.82^{+1.50}_{-1.30}$ & $2.52^{+0.12}_{-0.12}$ & $0.0744^{+0.0005}_{-0.0005}$ & 1898.7/1986 & 15.2  \\
&&&&&&&&&&&&\\
n10 & $1.81^{+0.10}_{-0.07}$ &$0.99^{+0.03}_{-0.03}$  & $0.30_{-0.18}^{+0.16}$ & $10.2^{+4.6}_{-4.4}$ & $6.506^{+0.184}_{-0.135}$ & $452^{+254}_{-130}$ & 120$_{-53}^{+57}$& $3.89_{-2.69}^{+1.63}$ & $2.45^{+0.27}_{-0.14}$ & $0.0744^{+0.0005}_{-0.0005}$ & 1904.1/1986 & 9.1  \\
&&&&&&&&&&&&\\
Ph & $1.82^{+0.11}_{-0.10}$ & $0.93^{+0.03}_{-0.03}$ & $0.31_{-0.17}^{+0.20}$ & $9.2^{+4.6}_{-4.3}$ & $6.394^{+0.094}_{-0.087}$ & $313^{+113}_{-91}$ & 88$^{+53}_{-46}$& $3.88_{-1.23}^{+1.34}$ & $2.86^{+0.27}_{-0.14}$ & $0.0735^{+0.0008}_{-0.0008}$ & 1906.5/1986 &  6.7 \\
\noalign{\smallskip}
\hline
    \end{tabular}
\end{subtable}\par
\bigskip
\begin{subtable}{\linewidth}
\centering\caption{\textit{RR models}: Col. I: Model name; Col. II: Photon index; Col III: Column density of the partial covering component; Col IV: Ionization factor of the partial covering component; Col. V: Observed redshift of the ionized absorber; Col VI: Covering factor; Col. VII: Width of the narrow Gaussian line; Col. VIII: Equivalent width of the narrow Gaussian line; Col. IX: Reflection fraction parameter; Col. X: Inner radius of the accretion disk; Col XI: Ionization factor of the relativistic reflection component; Col. XII: Cash statistic value/degrees of freedom; Col. XIII: Akaike Information Criterion.}
\begin{tabular}{lccccccccccccc}
\hline
\noalign{\smallskip}
\hline
\noalign{\smallskip}
Model & $\Gamma$ & $N_{H,pc}$&$\log\xi$ & $z_{obs}$ & $C_f$ & $\sigma_b$ & EW & R & r$_{in}$ & $\log\xi$  & C/dof & $\Delta$AIC \\
\noalign{\smallskip}
 & & ($10^{23}$\,cm$^{-2}$) & & & &(eV)  & (eV) &  & (r$_g$)  & & \\
\hline\noalign{\smallskip}
RR-I & $1.83^{+0.05}_{-0.04}$ &$1.51^{+0.37}_{-0.37}$ & & & $0.34^{+0.06}_{-0.03}$ & 10.2$^{+4.9}_{-4.6}$ &29$^{+24}_{-10}$& $0.17^{+0.10}_{-0.03}$ & $\geq$27& $2.82^{+0.22}_{-0.13}$ &   1921.2/1988 & \\
&&&&&&&&&&&&\\
RR-n8 & $1.72^{+0.02}_{-0.04}$ &$3.67^{+1.52}_{-1.07}$ &2.53$_{-0.08}^{+0.08}$ & 0.0744$_{-0.0005}^{+0.0005}$& $0.24^{+0.10}_{-0.12}$ & $10.8^{+4.8}_{-4.3}$ &28$^{+12}_{-10}$& $0.28^{+0.08}_{-0.06}$ & 53$^{+73}_{-27}$& $3.15^{+0.20}_{-0.22}$ &   1906.1/1986 & 11.1 \\
\noalign{\smallskip}
\hline
\end{tabular}
\end{subtable}
\tablefoot{model I: TBabs(zTBabs$\times$constant(zpl+zgauss)+((1+constant)$\times$(gsmooth$\times$MTL+constant$\times$MTS)+constant(zpl+zgauss)); \\
n8: TBabs(XSTARn8$\times$constant(zpl+zgauss)+((1+constant)$\times$(gsmooth$\times$MYTL+MYTS))+constant(zpl+zgauss)); \\
n10: same as n8 with a different XSTAR table (see Section~\ref{abs}); \\Ph: TBabs(phase(zpl+zgauss)+(1+constant)$\times$(gsmooth$\times$MYTL+MYTS)+constant(zpl+zgauss));\\ RR-I: TBabs$\times$zTBabs(relxill+zfeklor)+constant$\times$TBabs$\times$relxill;\\
RR-n8: TBabs$\times$XSTARn8(relxill+zfeklor)+constant$\times$TBabs$\times$relxill. 
}
\end{table*}

\subsection{An infalling partial absorber?}
\label{abs}

\begin{figure}
\centering
\includegraphics[scale=0.36]{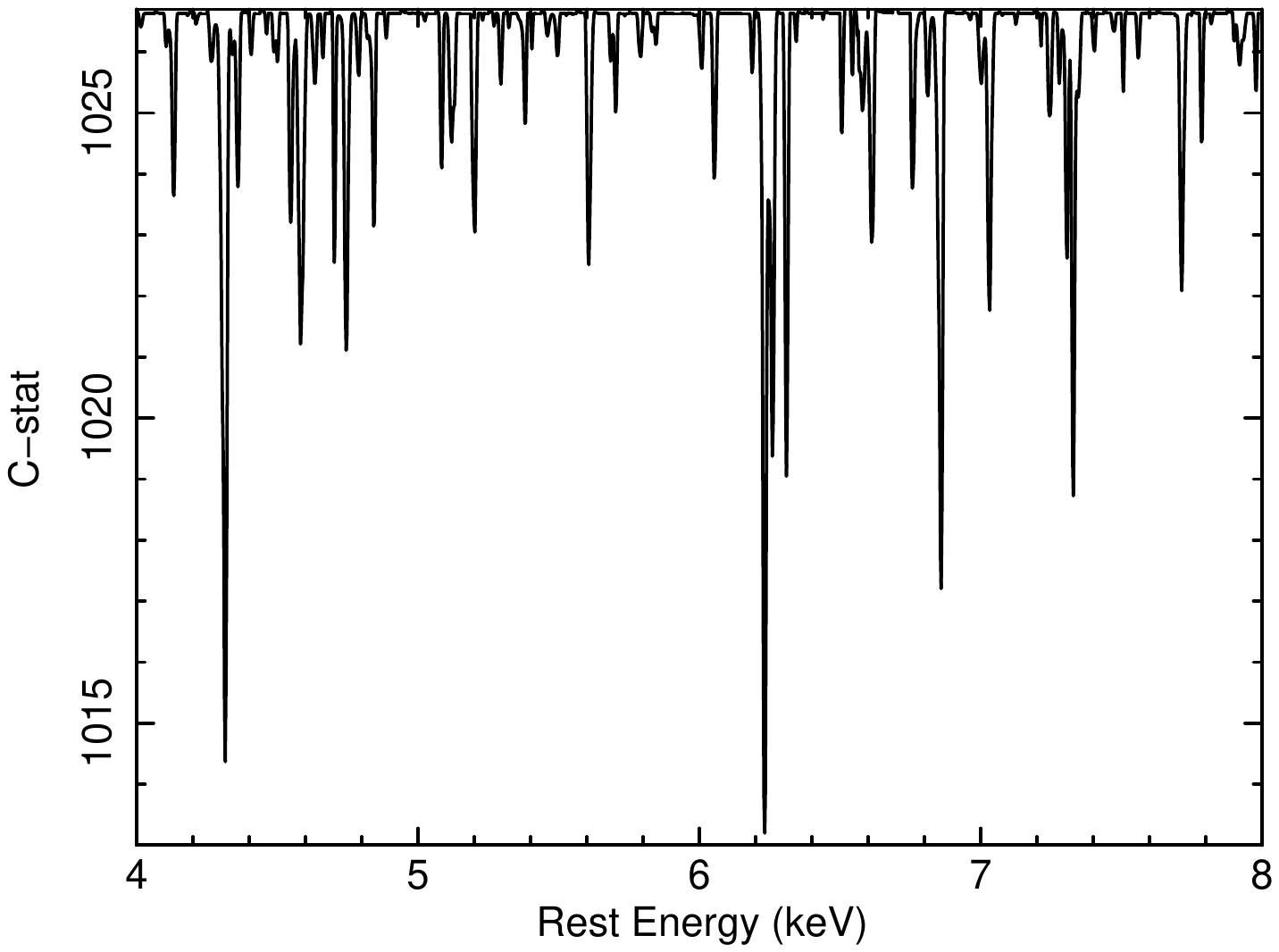}
\includegraphics[scale=0.36]{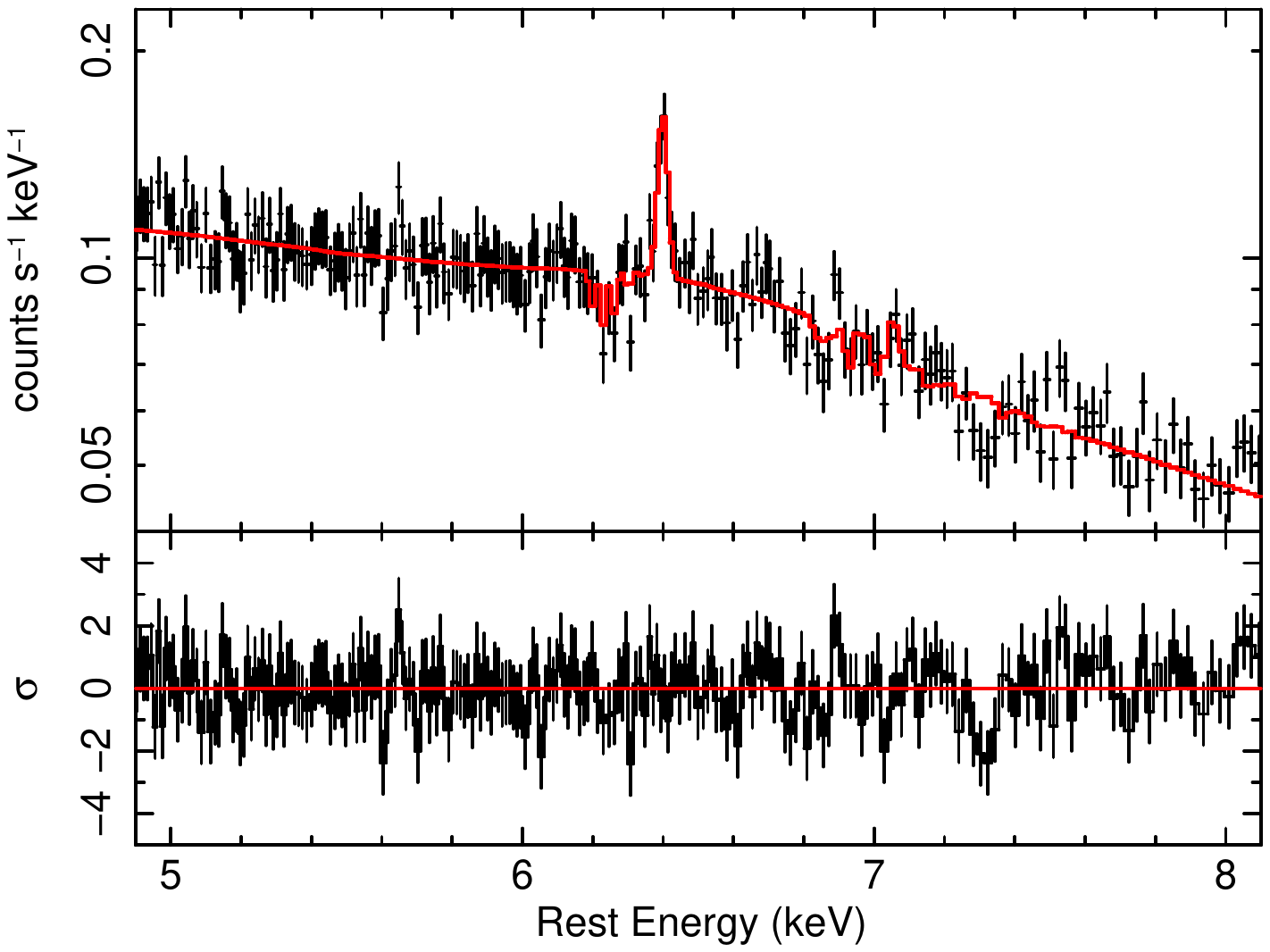}
\caption{\textit{Upper panel}: Confidence contours (rest frame energies) for the blind search for possible absorption features (see Section~\ref{abs}).
\textit{Lower panel}: Zoom-in of the spectral range where the strongest feature (absorption line at E=$6.230^{+0.005}_{-0.020}$\,keV) is measured.}
\label{ratabga}
\end{figure}

As described in the previous sections, a partial absorber of the primary emission is able to well model the data both in the PC and RR scenarios. As shown in Fig.~\ref{zpc} (here the PC scenario is presented, but the same is happening on the RR case), some spectral features remain visible in the spectrum of Mrk~509. In particular, possible absorption features can be identified at $E\sim4.2$, 6, 7, and possibly 8--9\,keV. When all these features are modeled with a Gaussian absorption line of fixed width $\sigma=5$\,eV, the statistically strongest line is found at E=$6.230^{+0.005}_{-0.020}$\,keV (see Fig.~\ref{ratabga} for the PC case). This addition improves the fit by $\Delta C\sim13$ in both scenarios.

To assess the importance of this absorption feature, we employed the Small Sample Akaike Information Criterion \citep[AICc,][]{akaike}. Specifically, we computed the change in AIC ($\Delta$AIC) when the line is removed from the model. A $\Delta$AIC below $-2$ provides strong evidence that the feature is required, while a value below $-10$ would indicate it is essential to explain the data. In our case, $\Delta$AIC$\sim$-9, supporting the presence of this absorption line. Furthermore, a closer inspection of the $\sim$6\,keV region suggests a more complex structure, with possible multiple dips between 6.0 and 6.1\,keV rest frame (see lower panel of Fig.~\ref{ratabga}).

\begin{figure}
\centering
\includegraphics[width=9.5cm, height=7cm]{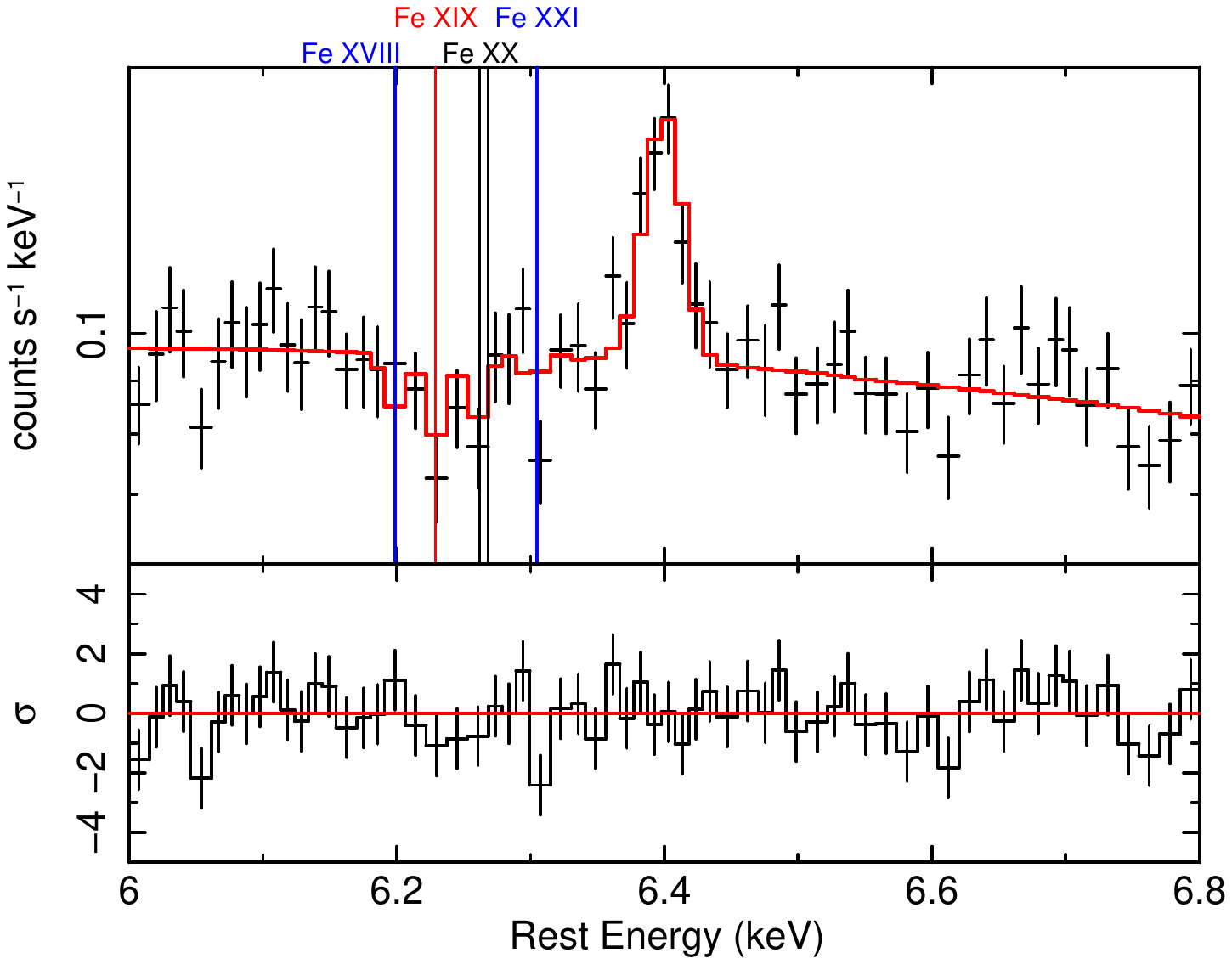}
\caption{6.0--6.8\,keV zoom-in of the spectral range where the strongest feature due to the ionized gas are measured (model n8 in Tab.~\ref{pcmy}).
The features measured in the 6.1--6.4\,keV are identified as due to Fe\,\textsc{xviii}--Fe\,\textsc{xxi}.
}

\label{infall}
\end{figure}

Given the apparent complexity of the absorption feature, we used photoionized absorber models as a partial covering component to fit the absorption feature at $\sim6$\,keV. We created grid tables using the \textsc{Xstar} code \citep[][]{xstar} setting an ionizing luminosity $L_{1-1000\,\mathrm{Ry}} \approx 3.1\times10^{45}$\,erg\,s$^{-1}$ (see Section~\ref{xrism}) and, following \citet{petrucci}, a SED dominated by a warm- plus a hot-corona component tuned on the \textit{XMM-Newton} + \textit{NuSTAR} data (Missaglia et al. in prep.).

As a first step, we focused on the PC case and assumed for the intervening gas a density of $n=10^8$\,cm$^{-3}$, consistent with the absorber being located at sub-pc scales, comparable to the BLR. This is, for instance, supported by the unified stratification models of AGN winds \citep[e.g.,][]{tombesi2013, gofford, reeves18, serafinelli} and density diagnostics using metastable lines in similar Seyfert galaxies which typically yield $n \geq 10^{8}$\,cm$^{-3}$ for the high-ionization phases \citep[e.g.,][]{mao}. Moreover, following the estimate obtained in Section~\ref{xrism} for the PC case, we tested also the possibility to have $n\sim10^{10}$\,cm$^{-3}$, thus assuming that the matter responsible for the emission of the broad Fe line is somehow linked to that responsible for the partial absorption. In both scenarios, we fixed the turbulence of the matter to be $v_{turb}=100$\,km\,s$^{-1}$ since the feature seems to be narrow (see Fig.~\ref{ratabga}). Models n8 and n10 and RR-n8 in Tab.~\ref{pcmy} report the results obtained using these \textsc{Xstar} grids.  The use of these models significantly  improved the quality of the fit in both scenarios, and provided a best fit result with absorbing gas density of $n=10^8$\,cm$^{-3}$, and log~$\xi\sim2.5$ (see Table~\ref{pcmy} models n8 and RR-n8, and Fig. \ref{infall}). 

 As a consistency check, we also tested a different photoionized model, namely PHASE model, based on \textsc{Cloudy} \citep[][]{cloudy}, which is designed to reproduce absorption features due to an ionized plasma with a geometry such that the central source emits an ionizing continuum with clouds of gas intercepting the line of sight (model Ph in the upper part of Tab. \ref{pcmy}). Despite the poorer statistical modeling of the data when compared with what obtained with the \textsc{Xstar} table, the PHASE modeling return consistent properties of the absorbing gas. 

\begin{figure}
\centering
\includegraphics[scale=0.35]{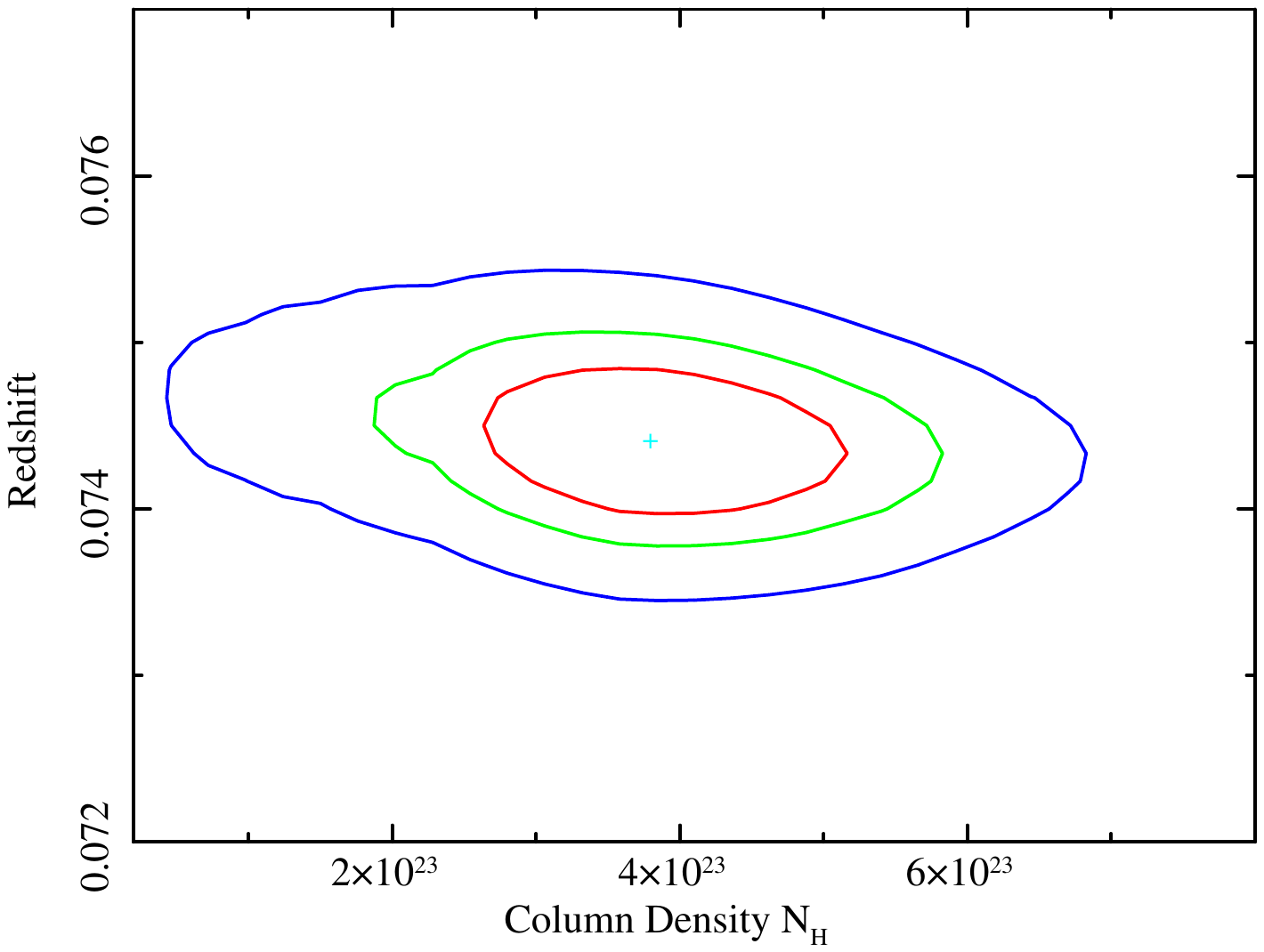}
\caption{\textit{Upper panel}: 99, 90, 68\% confidence contours for the $z$ vs $N_H$ parameters obatined using model n8 in Tab. \ref{pcmy}.
}
\label{infall2}
\end{figure}

 Interestingly, by anchoring the models to the absorption features observed between $\sim6$--$7$\,keV (see Fig.~\ref{infall2}), all the above models imply that the matter responsible for the ionized absorption is measured at a redshift $z_{\rm obs}\sim0.074$ (see lower panel of Fig. \ref{infall2}). Since this is larger than the systemic redshift of Mrk~509 ($z_{\rm sou}=0.034\pm0.001$), this finding—if confirmed—could indicate that the absorbing matter is infalling toward the central SMBH. In this energy range, the lines are primarily due to Fe\,\textsc{xviii}--Fe\,\textsc{xxi} transitions. 

\begin{figure}
\centering
\includegraphics[scale=0.37]{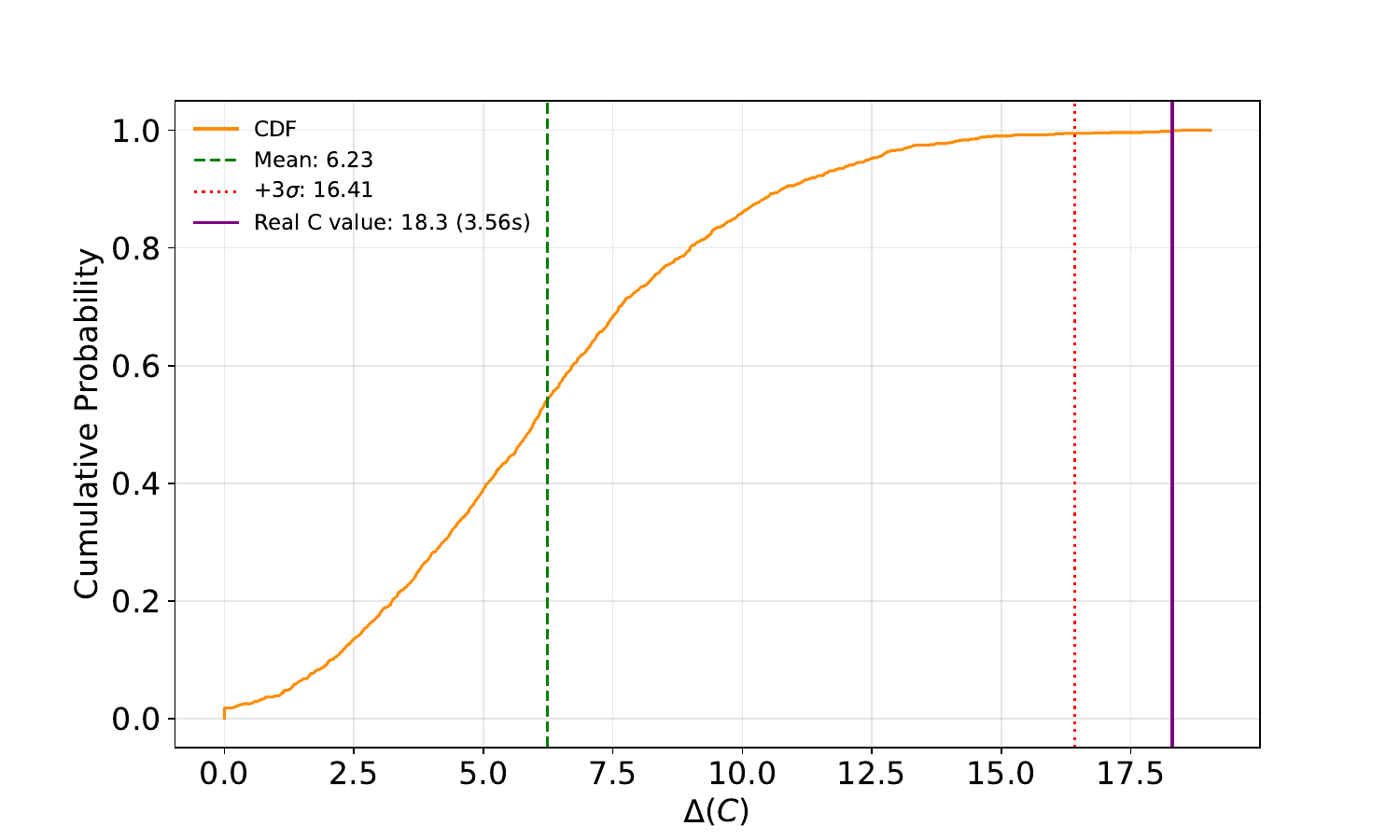}
\caption{Cumulative distribution of the $\Delta C$ statistic derived from 1500 Monte Carlo simulations, assuming Model~I (see Table~\ref{BB}) as the null hypothesis. Synthetic \textit{XRISM}/Resolve spectra were generated matching the exposure time of the real observation. For each simulation, $\Delta C$ is defined as the difference in C-statistic between the best fits obtained modeling the simulated datasets using Model~I and Model~n8 in Table~\ref{BB}. The vertical green line indicates the mean simulated $\Delta C$, while the red line marks the 3$\sigma$ confidence threshold. The purple line represents the $\Delta C$ value observed in the real data.
}
\label{cumu}
\end{figure}

To further test the significance of the infalling scenario, we performed Monte Carlo simulations under the PC scenario. Starting from the partial covering model (model I) presented in Table~\ref{pcmy}, we performed a suite of 1,500 simulations of \textit{XRISM}/Resolve spectra. These simulations were conducted using the rmf and arf generated for the data analysis, as previously described. To maintain consistency with the observational constraints, each simulated dataset was assigned an exposure time identical to that of the original observation. Subsequently, each of the 1500 simulated spectra was fitted using both the baseline Model I and the alternative model n8. For the model n8 fits, the ionization parameter was allowed to vary within the range of log($\xi/\mathrm{erg\,cm\,s}^{-1}$)=1.5--4.0, while the observed redshift was free to fluctuate between $z$=-0.2 and $z$=0.2. To ensure that each single fit did not stack in local minima, we forced the procedure to search for the errors on all free parameters multiple times before and after running the steppar command on the ionization parameter, the redshift, the column density of the ionized absorber and the continuum free parameters. For every  simulated dataset, we recorded the value of C-statistic for both models to calculate the difference, $\Delta(C)$, which provides the statistical basis for our cumulative distribution analysis presented in Fig.~\ref{cumu}. We found that, according to this test, the significance of our results is $\sim3.6\sigma$.

The current model leaves some residuals at a rest-frame energy of E$\sim$7.3\,keV (see Fig.~\ref{ratabga}). Adding a simple Gaussian absorption line yields a marginal improvement to the fit ($\Delta C\sim14$ for three additional free parameters). Nevertheless, the centroid energy of this feature (E$\sim$7.31\,keV, $\sigma\sim40$\,eV, EW$\sim$14\,eV) is perfectly consistent with previous findings \citep{ponti09}. If identified with highly ionized iron such as Fe~XXVI this absorption feature would indicate the presence of an outflowing wind with a velocity of the order of 13500 km/s.

It is interesting to note that completely consistent results regarding the partial absorber characteristics are obtained when using the RR baseline modeling instead of the PC scenario. As reported in the lower part of Table~\ref{pcmy}, utilizing an ionized absorber rather than a cold one improves the fit significantly ($\Delta C=15$ for two additional free parameters). More importantly, we obtain exactly the same properties for the absorber in terms of both its ionization state and velocity.

Finally, we tested the two scenarios by incorporating the \textit{NuSTAR} data for either Obs.1 (which was performed within the Resolve pointin, see Fig. \ref{timelog} pointing) and for Obs.1+Obs2 in the 10--60\,keV range. From a statistical point of view, the two models perform identically, and the resulting parameters are fully consistent with those obtained using the Resolve data alone. The only notable difference lies in the cross-calibration constant between \textit{XRISM}/Resolve and \textit{NuSTAR}. In the PC case, we obtain a cross-calibration constant of c$_r =1.08\pm0.05$, which is well in agreement with previous findings \citep{cross}. Conversely, in the RR case, we measure a higher constant of c$_r = 1.21\pm0.10$

\section{Discussion}

The \textit{XRISM}/Resolve spectrum of Mrk~509 displays the signatures of a complex structure at the energies of the Fe K lines. In particular, we have detected:
\begin{itemize}
    \item a narrow Fe K$\alpha$ component, most likely arising from the putative torus;
    \item a relativistic/broad Fe K$\alpha$ emission line originating from a region close to the central SMBH;
    \item spectral features consistent with a partial absorber, potentially associated with an infalling flow, though this interpretation remains model-dependent.
\end{itemize}

We estimate the dust sublimation radius for Mrk~509 to be $R_{\rm sub} \approx 0.74$\,pc, derived using the scaling relation in \citet{barvainis1987} for graphite grains with a sublimation temperature of $T_{\rm sub} \approx 1500$\,K and a bolometric luminosity of $L_{\rm bol} \approx 3.2 \times 10^{45}$\,erg\,s$^{-1}$ \citep{kaastra, detmers}.
In comparison, the measured width of the narrow Fe~K$\alpha$ core ($\sigma_E \approx 10$\,eV) corresponds to a virial emission radius of $R_{\rm FeK} \approx 2.0$\,pc, assuming a black hole mass of $M_{\rm BH} \sim 1.02 \times 10^8\,M_{\odot}$ \citep{li2024}.
This places the bulk of the neutral iron emission outside the sublimation boundary ($R_{\rm FeK} \sim 2.7 R_{\rm sub}$), indicating that the fluorescent reflector is located within the body of the dusty torus rather than at its innermost wall. This geometry is consistent with clumpy torus scenarios where X-ray fluorescence arises from cold, dusty clouds that are sufficiently shielded from the central engine \citep{ramos2017}.

As presented in Section \ref{xrism}, the best modeling of the data is obtained using a broad Fe emission detected at $E \approx 6.5 \pm 0.15$\,keV, with a width of $\sigma \approx 450 \pm 150$\,eV implies a velocity dispersion of $\sigma_v \approx 21\,000$\,km\,s$^{-1}$. Adopting the virial assumption and the same black hole mass, we estimate an emission radius of $R_{\rm broad} \approx 0.001$\,pc ($\sim 30$--$120\,r_g$). This places the production region roughly three orders of magnitude closer to the central engine than the dusty torus, consistent with the typical spatial scales of the BLR or the outer accretion disk. These spectral parameters are statistically consistent with those reported by \citet{ponti} based on long \textit{XMM-Newton} observations ($E=6.43\pm0.01$\,keV and $\sigma=0.22\pm0.05$\,keV). Notably, this physical scale is perfectly corroborated by the scenario in which the broad line originates from relativistic reflection of the accretion disk; when explicitly modeled as disk reflection, we derive a lower limit for the inner emission radius of r$\geq$27\,r$_g$ in excellent agreement with the virial estimate.

Our results for Mrk~509 align remarkably well with the emerging picture of the central engine revealed by recent \textit{XRISM} observations of other Type-1 AGN.
In NGC~4151, the \textit{XRISM} Collaboration \citep{XRISM4151} identified an intermediate Fe~K$\alpha$ component originating from the outer accretion disk or the BLR transition region ($r \sim 50$--$500\,r_g$). This spatial scale is directly comparable to the $r \sim 30$--$120\,r_g$ region we derive for Mrk~509. \citet{XRISM4151} interpret this feature as potentially arising from a warp in the accretion disk—which would locally enhance the illuminated solid angle—or from the base of a disk wind. It is worth noting that a similar wind scenario has recently been proposed for NGC~1068 \citep{graftonwaters21, bianchi2026}.

This consistency extends to other sources. In NGC~3783, \citet{mehdipour24} found that the broad Fe~K emission is significantly narrower than expected for reflection from the innermost stable circular orbit ($r \leq 10\,r_g$), favoring instead an origin at larger radii ($r \geq 20$--$50\,r_g$, even if it is worth noting that, subsequently, this result has been challenged by the relativistic modeling of this feature by \cite{li26}. More recently, NGC~7213 was shown by \citet{kammoun25} to exhibit a clear radial stratification. The authors resolved the Fe~K complex into a narrow core and a broad, asymmetric component that mirrors the optical H$\alpha$ profile, placing its origin at $r \sim 100\,r_g$. They interpret this as emission from the outer accretion disk, bridging the gap between the compact corona and the optical BLR.

These studies suggest that the canonical ``broad'' iron line—typically associated with extreme relativistic reflection from the inner few gravitational radii (e.g., as in MCG-6-30-15; \citealt{brenneman25})—is not always clearly detectable in nearby Seyfert galaxies. Instead, emission components originating from a transition zone at r$\sim 50-100\,r_g$ are possibly measured. This region, which may connect the turbulent accretion disk to the launch site of the BLR winds, can be an ubiquitous feature of Seyfert galaxies that microcalorimeters are now able to resolve.

The X-rays spectral analysis of Mrk~509 reveals that, independently of the specific ionization state of the absorbing matter, the modeling consistently requires a covering factor of $C_f \sim 30\%$ (see Tab.~\ref{pcmy}). To interpret the physical implications of this obscuration, we assume a slab geometry for the primary X-ray continuum, consistent with recent X-ray polarimetric findings from \textit{IXPE} \citep[e.g.,][]{gianolli23,pal,marinucci} and with the theoretical framework of a disk-corona system as proposed by \citet{haardt93} and \citet{haardt94}.

Assuming a typical coronal radius of $R_{\rm src} = 10 \, r_{\rm g}$ \citep{risaliti07, chartas09, fabian09, wilkins11} for a supermassive black hole with the same mass of Mrk~509, we have $r_{\rm g} \approx 1.5 \times 10^{13}$\,cm, thus the physical source radius of the corona is $R_{\rm src} \approx 1.5 \times 10^{14}$\,cm. To reproduce the observed covering factor of $\sim 30\%$ (interpreting $C_f$ as the area ratio), the individual cloudlet or obscuring structure must have a characteristic transverse diameter of $D_{\rm c} \approx 1.6 \times 10^{14}$\,cm (i.e. $\approx 10\,r_g$, a value consistent with what estimated for the ``bullets'' in PDS\,456, \citealt{pds456}). Combining the measured column density of the absorber, $N_{\rm H} \sim 1$--$4 \times 10^{23}$\,cm$^{-2}$, with these physical dimensions, and assuming a simple spherical or shell-like geometry where the path length is comparable to the diameter ($L \approx D_{\rm c}$), we estimate a volume density of $n \approx N_{\rm H}/D_{\rm c} \approx 1$--$7 \times 10^9$\,cm$^{-3}$ (taking into account the 90\% errors in the absorber column density and the covering factor).

However, this estimate assumes a uniform sphere and a central line of sight. When accounting for off-center sightlines (intercepting a chord rather than the full diameter) and the expected density stratification within the gas, this average value becomes a strict lower limit. The actual local density within the core of the absorber is expected to be significantly higher, likely on the order of $n \sim 10^{10}$\,cm$^{-3}$.

We can attempt to independently constrain these physical parameters by considering the ionization state of the absorber, bearing in mind the uncertainties in the detection. The data favor the presence of an ionized phase ($\log(\xi/\mathrm{erg\,cm\,s}^{-1})\sim2.5$, see Table~\ref{pcmy}) which, in the best-fit model, appears to possess a significant inflow velocity of $v_{in}\approx11\,000$\,km\,s$^{-1}$. If we assume that the gas dynamics are solely governed by the gravitational potential of the central SMBH, this free-fall velocity would place the material at a distance of $r \approx 1500\,r_{\rm g}$ ($\approx 2.2 \times 10^{16}$\,cm). Using this distance and the observed ionization parameter, we estimate a local gas density of $n \approx 2 \times 10^{10}$\,cm$^{-3}$.   
When taking into account the outward radiation pressure, the effective gravitational acceleration is reduced by a factor of ($1-\Gamma_{rad}$) \citep[e.g.,][]{proga, king}, where $\Gamma_{rad}\sim$0.25 is the Eddington ratio for Mrk~509, yielding to  $r \approx 1150\,r_{\rm g}$ and $n \approx 3.5 \times 10^{10}$\,cm$^{-3}$).

This value is remarkably consistent with the lower limit derived from the geometrical constraints described above. This tentatively suggests a scenario where the absorber might consist of dense, BLR-like material located within the inner few thousand gravitational radii. In this picture, the gas participates in a non-Keplerian inflow, fragmented into blobs rather than forming a smooth, continuous flow, consistent with failed wind clumps ``raining'' back toward the central engine \citep[e.g.,][]{ghisellini,proga,giustini19}. It is worth noting that also detailed optical studies of the BLR properties of Mrk~509 highlighted the presence of clouds with radial motion components, though in outflow \citep[][]{gravity}.

It is worth noting a tension between the spectral modeling results and the physical estimates derived above. Specifically, the fit statistically prefers model~n8 ($n=10^8 cm^{-3}$) over model~n10 (see Table~\ref{pcmy}), even though the latter employs a density ($n=10^{10}\,cm^{-3}$) that is consistent with the geometrical constraints and the virial requirements for the inflow. First of all it is important to stress that the statistical difference between these two solutions is not large ($\Delta$C=5.2), suggesting that the data do not rule out the high-density scenario on statistical grounds alone. On the other hand this apparent discrepancy can be reconciled assuming that the absorber possesses a 'cometary' structure \citep[][]{maiolino10,risaliti2011}. In this scenario, the absorber relies on a dense, compact core. As this core plunges inward, the intense radiation field ablates its outer layers, generating a diffuse, elongated tail with low density material. The radiation pressure directs this tail outward, effectively pointing it toward the observer. Consequently, our line of sight passes through the full length of this lower-density tail before interacting with the core. The spectral modeling is therefore weighted toward the physical conditions of the extended tail, while the measured column density ($N_H \sim 10^{23} cm^{-2}$) results from integrating along this extended path length (L$\sim10^{15}$ cm), consistent with the proposed geometry.

The energetic implications of these potential ``raining'' clumps/comets are non-negligible. Traveling at a free-fall velocity of $v_{\mathrm{in}} \approx 11\,000$\,km\,s$^{-1}$, a single clump would carry significant kinetic energy. Adopting the derived physical diameter of $D_{\mathrm{c}} \approx 1.6 \times 10^{14}$\,cm (implying a spherical volume of $V \approx 2.1 \times 10^{42}$\,cm$^3$) and a fiducial density of $n=10^{10}$\,cm$^{-3}$, we estimate a single clump mass of $M_{\mathrm{c}} \approx 3.6 \times 10^{28}$\,g (assuming $M_c \approx n V m_p$). Consequently, the kinetic energy associated with this structure would be $K = \frac{1}{2} M_{\mathrm{c}} v_{\mathrm{in}}^2 \approx 2 \times 10^{46}$\,erg. 

Although the kinetic energy carried by a single cloudlet may be substantial, it represents a transient energy injection. Given the average bolometric luminosity of the source ($L_{\rm bol} \sim 3 \times 10^{45}\rm \,erg\, s^{-1}$) the energy released by the accretion of a single clump would sustain the AGN output for only a few seconds (assuming efficient thermalization). Therefore, in order for this "raining" mechanism to have a significant, sustained impact on the global energy budget of the system, such accretion events cannot be sporadic. On the other hand, these cloudlets impacting the disk may play a crucial dynamical role beyond merely replenishing the fuel supply. The collision of such massive, dense clumps with the accretion disk may induce local instabilities and density perturbations within the flow. These stochastic inhomogeneities can then propagate inward through the disk on viscous timescales, effectively acting as the physical "seeds" for the propagating fluctuations model \citep[e.g.,][]{lyubarskii97, arevalo06}. In this scenario, the raining material may contribute to the intrinsic variability of the accretion flow, providing a mechanism to explain the observed broadband timing properties, including the characteristic time lags detected between different X-ray energy bands \citep[][]{demarco13}. 

It is finally worth mentioning here that the possible detection of a concurrent UFO signature in the form of an absorption line at E$\sim$7.3 keV \citep[see also][]{ponti09} is in agreement with the scenario depicted in \cite{elvis17} for the formation of BLR in AGNs.

\section{Conclusions}
In this work, we presented a detailed spectral analysis of the Seyfert~1 galaxy Mrk~509, focusing on the high-resolution X-ray spectrum obtained by the \textit{XRISM}/Resolve calorimeter, complemented by simultaneous \textit{NuSTAR} and \textit{XMM-Newton} observations. Our analysis leads to the following main conclusions:

\begin{itemize}
    \item The Fe K complex decomposition: Thanks to the unprecedented resolution of \textit{XRISM}, we successfully disentangled the various components of the Fe K complex. We identified a narrow neutral Fe~K$\alpha$ core ($\sigma < 10$\,eV), originating from distant material consistent with the dusty torus (as modeled by \texttt{MYTorus}). Additionally, a broad emission component is required. We modeled it using either a broad Gaussian in emission or a relativistic reflection obtaining that it must originate at $r \sim30-120\,r_g$ from the central engine, consistent with the BLR or the outer accretion disk, rather than the innermost regions.

    \item Partial covering vs. pure relativistic reflection: we tested two scenarios to explain the spectral curvature and the 2-10 keV complexity. While a pure relativistic line can statistically reproduce the curvature, it yields a primary photon index ($\Gamma \sim 1.5-1.6$, see Tab. \ref{BB}) that is quite flat when compared with average values for Seyfert galaxies. Conversely, a partial covering model (for both models n8 and RR-n8 in Tab.  \ref{pcmy}) retrieves a canonical photon index ($\Gamma \sim 1.7-1.8$).

    \item Potential indications of a ``raining'' absorber: The modeling suggests the presence of an ionized partial absorber ($\log(\xi/\mathrm{erg\,cm\,s}^{-1})\sim2.5$) which, under specific density assumptions ($n \approx 10^{8-10}$\,cm$^{-3}$), is consistent with being inflowing at a high velocity of $v_{\mathrm{in}} \approx 11\,000$\,km\,s$^{-1}$. If confirmed, the physical modeling places this absorber at $r \approx 1100-1500\,r_g$, consistent with a clumpy, fragmented flow—likely interpreted as ``failed wind'' clumps losing angular momentum and raining back toward the central black hole, possibly with a cometary-like shape.

\end{itemize}

Future monitoring with \textit{XRISM} and \textit{NewAthena} will be crucial to firmly confirm the existence and track the variability of these absorption features in Mrk~509 and in other sources. This is mandatory in order to firmly establish, if any, the possible connections between the infalling gas dynamics and the accretion flow that must fuel the central SMBH in AGN.

\begin{acknowledgements}

The authors are grateful to the anonymous referee for their careful reading of the manuscript and valuable feedback, which significantly strengthened this work. M.D., V.M., V.B., M.C., A.L., S.B., A.C., E.N., F.N., C.P., E.B. and A.T. acknowledge financial support from the Bando Ricerca Fondamentale INAF 2023, Large Program 1.05.23.01.06 (“The XRISM-to-XIFU (X2X) Agreement and Beyond: entering a new Era of High Resolution X-Ray Spectroscopy”). M.D, M.C. F.N and A.L. acknowledge financial support from PRIN MUR 2022 DRAGON 2022K9N5B4. E.B.  acknowledges the support of  the INAF GO grant ``A JWST/MIRI MIRACLE: Mid-IR Activity of Circumnuclear Line Emission'' and of the ``Ricerca Fondamentale 2024'' INAF program (mini-grant 1.05.24.07.01). P.O.P. acknowledges support from the CNES french spatial agency and from the « Action Thématique Phénomènes Extrêmes et Multimessager » of the Astronomy-Astrophysics National Programme from INSU/CNRS. C.P. is funded by INAF Large Grant 2023 BLOSSOM F.O. 1.05.23.01.13. R.S. acknowledges funding from the CAS-ANID grant number CAS220016.

\end{acknowledgements}



\begin{thebibliography}{}

\bibitem[Akaike(1974)]{akaike} Akaike, H.\ 1974, IEEE Transactions on Automatic Control, 19, 716. doi:10.1109/TAC.1974.1100705

\bibitem[Antonucci(1993)]{antonucci1993} Antonucci, R.\ 1993, \araa, 31, 473. doi:10.1146/annurev.aa.31.090193.002353

\bibitem[Ar{\'e}valo \& Uttley(2006)]{arevalo06} Ar{\'e}valo, P. \& Uttley, P.\ 2006, \mnras, 367, 801. doi:10.1111/j.1365-2966.2006.10133.x

\bibitem[Arnaud et al.(1999)]{xspec} Arnaud, K., Dorman, B., \& Gordon, C.\ 1999, Astrophysics Source Code Library. ascl:9910.005

\bibitem[Barvainis(1987)]{barvainis1987} Barvainis, R.\ 1987, \apj, 320, 537. doi:10.1086/165571

\bibitem[Bianchi et al.(2009)]{bianchi09} Bianchi, S., Guainazzi, M., Matt, G., et al.\ 2009, \aap, 495, 421

\bibitem[Bianchi et al.(2026)]{bianchi2026} Bianchi, S., Vander Meulen, B., Bertola, E., et al.\ 2026, arXiv:2602.16252. doi:10.48550/arXiv.2602.16252


\bibitem[Braito et al.(2018)]{braito} Braito, V., Reeves, J.~N., Matzeu, G.~A., et al.\ 2018, \mnras, 479, 3, 3592. doi:10.1093/mnras/sty1697

\bibitem[Brenneman et al.(2025)]{brenneman25} Brenneman, L.~W., Wilkins, D.~R., Ogorza{\l}ek, A., et al.\ 2025, \apj, 995, 2, 200. doi:10.3847/1538-4357/ae1225

\bibitem[Cappi(2006)]{cappi2006} Cappi, M.\ 2006, Astronomische Nachrichten, 327, 10, 1012. doi:10.1002/asna.200610639

\bibitem[Cappi et al.(2009)]{cappi2009} Cappi, M., Tombesi, F., Bianchi, S., et al.\ 2009, \aap, 504, 2, 401. doi:10.1051/0004-6361/200912137


\bibitem[Chartas et al.(2002)]{chartas} Chartas, G., Brandt, W.~N., Gallagher, S.~C., et al.\ 2002, \apj, 579, 1, 169. doi:10.1086/342744

\bibitem[Chartas et al.(2009)]{chartas09} Chartas, G., Kochanek, C.~S., Dai, X., et al.\ 2009, \apj, 693, 174. doi:10.1088/0004-637X/693/1/174

\bibitem[Chartas et al.(2021)]{chartas21} Chartas, G., Cappi, M., Vignali, C., et al.\ 2021, \apj, 920, 24. doi:10.3847/1538-4357/ac11f2

\bibitem[Collin et al.(2006)]{collin06}  Collin, S., Kawaguchi, T., Peterson, B.~M., \& Vestergaard, M.\ 2006, \aap, 456, 75. doi:10.1051/0004-6361:20064878

\bibitem[Dadina et al.(2005)]{dadina} Dadina, M., Cappi, M., Malaguti, G., et al.\ 2005, \aap, 442, 2, 461. doi:10.1051/0004-6361:20042487

\bibitem[Dadina(2007)]{dadina07} Dadina, M.\ 2007, \aap, 461, 3, 1209. doi:10.1051/0004-6361:20065734

\bibitem[Dadina(2008)]{dadina08} Dadina, M.\ 2008, \aap, 485, 417

\bibitem[Dadina et al.(2018)]{dadina18} Dadina, M., Vignali, C., Cappi, M., et al.\ 2018, \aap, 610, L13. doi:10.1051/0004-6361/201732386

\bibitem[Dauser et al.(2014)]{relxill} Dauser, T., Garcia, J., Parker, M.~L., et al.\ 2014, \mnras, 444, L100. doi:10.1093/mnrasl/slu125


\bibitem[De Marco et al.(2013)]{demarco13} De Marco, B., Ponti, G., Cappi, M., et al.\ 2013, \mnras, 431, 2441. doi:10.1093/mnras/stt339

\bibitem[Detmers et al.(2011)]{detmers} Detmers, R.~G., Kaastra, J.~S., Steenbrugge, K.~C., et al.\ 2011, \aap, 534, A38. doi:10.1051/0004-6361/201116899

\bibitem[Elvis(2017)]{elvis17} Elvis, M.\ 2017, \apj, 847, 56

\bibitem[Fabian et al.(2000)]{fabian2000} Fabian, A.~C., Iwasawa, K., Reynolds, C.~S., \& Young, A.~J.\ 2000, \pasp, 112, 1145

\bibitem[Fabian et al.(2009)]{fabian09} Fabian, A.~C., Zoghbi, A., Ross, R.~R., et al.\ 2009, \nat, 459, 540. doi:10.1038/nature08007

\bibitem[Ferland et al.(1998)]{cloudy} Ferland, G.~J., Korista, K.~T., Verner, D.~A., et al.\ 1998, \pasp, 110, 749, 761. doi:10.1086/316190

\bibitem[Garc{\'\i}a et al.(2016)]{garcia2016} Garc{\'\i}a, J.~A., Fabian, A.~C., Kallman, T.~R., et al.\ 2016, \mnras, 462, 1, 751. doi:10.1093/mnras/stw1696

\bibitem[Gaspari et al.(2020)]{gaspari} Gaspari, M., Tombesi, F., \& Cappi, M.\ 2020, Nat. Astron., 4, 10. doi:10.1038/s41550-019-0970-1

\bibitem[Ghisellini et al.(2004)]{ghisellini} Ghisellini, G., Haardt, F., \& Matt, G.\ 2004, \aap, 413, 535. doi:10.1051/0004-6361:20031562

\bibitem[Gianolli et al.(2023)]{gianolli23} Gianolli, V.~E., Kim, D.~E., Bianchi, S., et al.\ 2023, \mnras, 523, 3, 4468. doi:10.1093/mnras/stad1697

\bibitem[Gianolli et al.(2024)]{gianolli} Gianolli, V.~E., Bianchi, S., Petrucci, P.-O., et al.\ 2024, \aap, 687, A235. doi:10.1051/0004-6361/202348908

\bibitem[Giustini et al.(2017)]{giustini} Giustini, M., Costantini, E., De Marco, B., et al.\ 2017, \aap, 597, A66. doi:10.1051/0004-6361/201628686

\bibitem[Giustini \& Proga(2019)]{giustini19} Giustini, M. \& Proga, D.\ 2019, \aap, 630, A94. doi:10.1051/0004-6361/201833810

\bibitem[Gofford et al.(2013)]{gofford} Gofford, J., Reeves, J.~N., Tombesi, F., et al.\ 2013, \mnras, 430, 60. doi:10.1093/mnras/sts629

\bibitem[Grafton-Waters et al.(2021)]{graftonwaters21} Grafton-Waters, S., Branduardi-Raymont, G., Mehdipour, M., et al.\ 2021, \aap, 649, A120

\bibitem[GRAVITY Collaboration et al.(2024)]{gravity} GRAVITY Collaboration, Amorim, A., Bourdarot, G., et al.\ 2024, \aap, 684, A167. doi:10.1051/0004-6361/202348167

\bibitem[Haardt \& Maraschi(1993)]{haardt93} Haardt, F. \& Maraschi, L.\ 1993, \apj, 413, 507. doi:10.1086/173020

\bibitem[Haardt et al.(1994)]{haardt94} Haardt, F., Maraschi, L., \& Ghisellini, G.\ 1994, \apjl, 432, L95

\bibitem[Harrison et al.(2013)]{harrison} Harrison, F.~A., Craig, W.~W., Christensen, F.~E., et al.\ 2013, \apj, 770, 103. doi:10.1088/0004-637X/770/2/103

\bibitem[Heckman \& Best(2014)]{heckman} Heckman, T.~M. \& Best, P.~N.\ 2014, \araa, 52, 589. doi:10.1146/annurev-astro-081913-035722

\bibitem[H{\"o}lzer et al.(1997)]{holzer97} H{\"o}lzer, G., Fritsch, M., Deutsch, M., H{\"a}rtwig, J., \& F{\"o}rster, E.\ 1997, \pra, 56, 4554. doi:10.1103/PhysRevA.56.4554

\bibitem[Huchra et al.(1993)]{huchra} Huchra, J., Latham, D.~W., da Costa, L.~N., et al.\ 1993, \aj, 105, 1637. doi:10.1086/116543

\bibitem[Ishisaki et al.(2025)]{resolve} Ishisaki, Y., Kelley, R.~L., Awaki, H., et al.\ 2025, Journal of Astronomical Telescopes, Instruments, and Systems, 11, 042023. doi:10.1117/1.JATIS.11.4.042023


\bibitem[Jansen et al.(2001)]{jansen} Jansen, F., Lumb, D., Altieri, B., et al.\ 2001, \aap, 365, L1. doi:10.1051/0004-6361:20000036

\bibitem[Kaastra et al.(2012)]{kaastra} Kaastra, J.~S., Detmers, R.~G., Mehdipour, M., et al.\ 2012, \aap, 539, A117. doi:10.1051/0004-6361/201118161

\bibitem[Kallman \& Bautista(2001)]{xstar} Kallman, T. \& Bautista, M.\ 2001, \apjs, 133, 1, 221. doi:10.1086/319184

\bibitem[Kammoun et al.(2025)]{kammoun25} Kammoun, E., Kawamuro, T., Murakami, K., et al.\ 2025, arXiv:2510.24971

\bibitem[King \& Pounds(2015)]{king} King, A. \& Pounds, K.\ 2015, \araa, 53, 115. doi:10.1146/annurev-astro-082214-122316

\bibitem[Kormendy \& Ho(2013)]{kormendy} Kormendy, J. \& Ho, L.~C.\ 2013, \araa, 51, 1, 511. doi:10.1146/annurev-astro-082708-101811

\bibitem[Kumari et al.(2021)]{kumari} Kumari, N., Pal, M., Naik, S., et al.\ 2021, \pasa, Complex optical/UV and X-ray variability in Seyfert 1 galaxy Mrk 509, 38, e042. doi:10.1017/pasa.2021.41

\bibitem[Laha et al.(2021)]{laha} Laha, S., Reynolds, C.~S., Reeves, J., et al.\ 2021, Nature Astronomy, 5, 13. doi:10.1038/s41550-020-01255-2

\bibitem[Li et al.(2024)]{li2024} Li, Y.-R., Hu, C., Yao, Z.-H., et al.\ 2024, \apj, 974, 1, 86. doi:10.3847/1538-4357/ad6906

\bibitem[Li et al.(2026)]{li26} Li, C., Kaastra, J.~S., Gu, L., et al.\ 2026, \aap, 706, A255. doi:10.1051/0004-6361/202557710


\bibitem[Longinotti et al.(2007)]{longinotti} Longinotti, A.~L., Sim, S.~A., Nandra, K., et al.\ 2007, \mnras, 374, 1, 237. doi:10.1111/j.1365-2966.2006.11138.x

\bibitem[Lyubarskii(1997)]{lyubarskii97} Lyubarskii, Y.~E.\ 1997, \mnras, 292, 679. doi:10.1093/mnras/292.3.679

\bibitem[Maiolino et al.(2010)]{maiolino10} Maiolino, R., Risaliti, G., Salvati, M., et al.\ 2010, \aap, 517, A47. doi:10.1051/0004-6361/200913985

\bibitem[Mao et al.(2017)]{mao} Mao, J., Kaastra, J.~S., Mehdipour, M., et al.\ 2017, \aap, 607, A100. doi:10.1051/0004-6361/201731378

\bibitem[Marinucci et al.(2022)]{marinucci} Marinucci, A., Muleri, F., Dovciak, M., et al.\ 2022, \mnras, 516, 4, 5907. doi:10.1093/mnras/stac2634

\bibitem[Matzeu et al.(2016)]{matzeu16} Matzeu, G.~A., Reeves, J.~N., Nardini, E., et al.\ 2016, \mnras, 458, 1311. doi:10.1093/mnras/stw354

\bibitem[Matzeu et al.(2017)]{matzeu17} Matzeu, G.~A., Reeves, J.~N., Braito, V., et al.\ 2017, \mnras, 472, L15. doi:10.1093/mnrasl/slx129

\bibitem[Matzeu et al.(2023)]{matzeu23} Matzeu, G.~A., Brusa, M., Lanzuisi, G., et al.\ 2023, \aap, 672, A24. doi:10.1051/0004-6361/202244368

\bibitem[Mehdipour et al.(2011)]{mehdip2011} Mehdipour, M., Branduardi-Raymont, G., Kaastra, J.~S., et al.\ 2011, \aap, 534, A39. doi:10.1051/0004-6361/201116875

\bibitem[Mehdipour et al.(2025)]{mehdipour24} Mehdipour, M., Kaastra, J.~S., Eckart, M.~E., et al.\ 2025, \aap, 699, A228

\bibitem[Middei et al.(2019)]{middei19} Middei, R., Bianchi, S., Petrucci, P.~O., et al.\ 2019, \mnras, 483, 4695

\bibitem[Mizumoto et al.(2026)]{mizumoto26} Mizumoto, M., Reeves, J.~N., Braito, V., et al.\ 2026, \apj, 997, 2, 219

\bibitem[Murphy \& Yaqoob(2009)]{mytorus} Murphy, K.~D. \& Yaqoob, T.\ 2009, \mnras, 397, 1549. doi:10.1111/j.1365-2966.2009.15025.x

\bibitem[Nandra et al.(1997)]{nandra97}  Nandra, K., George, I.~M., Mushotzky, R.~F., Turner, T.~J., \& Yaqoob, T.\ 1997, \
\bibitem[Nandra \& Pounds(1994)]{nandra94} Nandra, K. \& Pounds, K.~A.\ 1994, \mnras, 268, 405

\bibitem[Nandra et al.(1999)]{nandra99} Nandra, K., George, I.~M., Mushotzky, R.~F., et al.\ 1999, \apjl, 523, L17. doi:10.1086/312261


\bibitem[Nandra et al.(2007)]{nandra07} Nandra, K., O'Neill, P.~M., George, I.~M., \& Reeves, J.~N.\ 2007, \mnras, 382, 194. doi:10.1111/j.1365-2966.2007.12331.x

\bibitem[Nardini et al.(2015)]{nardini} Nardini, E., Reeves, J.~N., Gofford, J., et al.\ 2015, Science, 347, 860. doi:10.1126/science.1259202

\bibitem[Negrete et al.(2012)]{negrete} Negrete, C.~A., Dultzin, D., Marziani, P., et al.\ 2012, \apj, 757, 1, 62. doi:10.1088/0004-637X/757/1/62

\bibitem[Pal et al.(2023)]{pal} Pal, I., Stalin, C.~S., Chatterjee, R., et al.\ 2023, Journal of Astrophysics and Astronomy, 44, 2, 87. doi:10.1007/s12036-023-09981-5

\bibitem[Peca et al.(2025)]{peca25} Peca, A., Koss, M.~J., Serafinelli, R., et al.\ 2025, \apj, 989, 1, 84. doi:10.3847/1538-4357/adea4a

\bibitem[Peterson et al.(2004)]{peterson04} Peterson, B.~M., Ferrarese, L., Gilbert, K.~M., et al.\ 2004, \apj, 613, 682. doi:10.1086/423269

\bibitem[Petrucci et al.(2013)]{petrucci2013} Petrucci, P.-O., Paltani, S., Malzac, J., et al.\ 2013, \aap, 549, A73. doi:10.1051/0004-6361/201219956


\bibitem[Petrucci et al.(2018)]{petrucci} Petrucci, P.-O., Ursini, F., De Rosa, A., et al.\ 2018, \aap, 611, A59. doi:10.1051/0004-6361/201731580

\bibitem[Piconcelli et al.(2005)]{piconcelli05} Piconcelli, E., Jimenez-Bail{\'o}n, E., Guainazzi, M., et al.\ 2005, \aap, 432, 15

\bibitem[Ponti et al.(2009)]{ponti09} Ponti, G., Cappi, M., Vignali, C., et al.\ 2009, \mnras, 394, 3, 1487. doi:10.1111/j.1365-2966.2009.14411.x

\bibitem[Ponti et al.(2013)]{ponti} Ponti, G., Cappi, M., Costantini, E., et al.\ 2013, \aap, 549, A72. doi:10.1051/0004-6361/201219450

\bibitem[Pounds et al.(2001)]{pounds} Pounds, K., Reeves, J., O'Brien, P., et al.\ 2001, \apj, 559, 1, 181. doi:10.1086/322315

\bibitem[Pounds et al.(2003)]{pounds03} Pounds, K.~A., Reeves, J.~N., King, A.~R., et al.\ 2003, \mnras, 345, 3, 705. doi:10.1046/j.1365-8711.2003.07006.x

\bibitem[Proga \& Kallman(2004)]{proga} Proga, D. \& Kallman, T.~R.\ 2004, \apj, 616, 2, 688. doi:10.1086/425117

\bibitem[Ramos Almeida \& Ricci(2017)]{ramos2017} Ramos Almeida, C. \& Ricci, C.\ 2017, Nature Astronomy, 1, 679. doi:10.1038/s41550-017-0232-z

\bibitem[Reeves et al.(2002)]{reeves02} Reeves, J.~N., Wynn, G., O'Brien, P.~T., et al.\ 2002, \mnras, 335, 126. doi:10.1046/j.1365-8711.2002.05612.x

\bibitem[Reeves et al.(2009)]{reeves09} Reeves, J.~N., O'Brien, P.~T., Braito, V., et al.\ 2009, \apj, 701, 493. doi:10.1088/0004-637X/701/1/493

\bibitem[Reeves et al.(2014)]{reeves14} Reeves, J.~N., Braito, V., Gofford, J., et al.\ 2014, \apj, 780, 45. doi:10.1088/0004-637X/780/1/45

\bibitem[Reeves et al.(2018)]{reeves18} Reeves, J.~N., Braito, V., Nardini, E., et al.\ 2018, \apjl, 854, L8. doi:10.3847/2041-8213/aaa8df

\bibitem[Ricci et al.(2017)]{ricci17} Ricci, C., Trakhtenbrot, B., Koss, M.~J., et al.\ 2017, \apjs, 233, 17

\bibitem[Risaliti et al.(2007)]{risaliti07} Risaliti, G., Elvis, M., Fabbiano, G., et al.\ 2007, \apjl, 659, L111. doi:10.1086/517881

\bibitem[Risaliti et al.(2011)]{risaliti2011} Risaliti, G., Nardini, E., Salvati, M., et al.\ 2011, \mnras, 410, 2, 1027. doi:10.1111/j.1365-2966.2010.17503.x


\bibitem[Ross \& Fabian(1993)]{ross1993} Ross, R.~R. \& Fabian, A.~C.\ 1993, \mnras, 261, 74. doi:10.1093/mnras/261.1.74

\bibitem[Serafinelli et al.(2019)]{serafinelli} Serafinelli, R., Tombesi, F., Vagnetti, F., et al.\ 2019, \aap, 627, A121

\bibitem[Shakura \& Sunyaev(1973)]{shakura1973} Shakura, N.~I. \& Sunyaev, R.~A.\ 1973, \aap, 24, 337. 

\bibitem[{\'S}niegowska et al.(2021)]{sniegowska} {\'S}niegowska, M., Marziani, P., Czerny, B., et al.\ 2021, \apj, 910, 2, 115. doi:10.3847/1538-4357/abe1c8

\bibitem[Tanaka et al.(1995)]{tanaka} Tanaka, Y., Nandra, K., Fabian, A.~C., et al.\ 1995, \nat, 375, 6533, 659. doi:10.1038/375659a0

\bibitem[Tashiro et al.(2024)]{xrism} Tashiro, M., Watanabe, S., Maejima, H., et al.\ 2024, \procspie, 13093, 130931G. doi:10.1117/12.3019325

\bibitem[Tashiro et al.(2025)]{xrism2025} Tashiro, M., Kelley, R., Watanabe, S., et al.\ 2025, \pasj, 77, S1. doi:10.1093/pasj/psaf023


\bibitem[Tombesi et al.(2010)]{tombesi10} Tombesi, F., Cappi, M., Reeves, J.~N., et al.\ 2010, \aap, 521, A57

\bibitem[Tombesi et al.(2011)]{tombesi} Tombesi, F., Cappi, M., Reeves, J.~N., et al.\ 2011, \apj, 742, 1, 44. doi:10.1088/0004-637X/742/1/44

\bibitem[Tombesi et al.(2013)]{tombesi2013} Tombesi, F., Cappi, M., Reeves, J.~N., et al.\ 2013, \mnras, 430, 2, 1102. doi:10.1093/mnras/sts692

\bibitem[Wilkins \& Fabian(2011)]{wilkins11} Wilkins, D.~R. \& Fabian, A.~C.\ 2011, \mnras, 414, 1269. doi:10.1111/j.1365-2966.2011.18458.x


\bibitem[XRISM Collaboration(2025)]{pds456} Xrism Collaboration, Audard, M., Awaki, H., et al.\ 2025, \nat, 641, 8065, 1132. doi:10.1038/s41586-025-08968-2

\bibitem[XRISM Collaboration et al.(2024)]{XRISM4151} 
{XRISM Collaboration}, Audard, M., Awaki, H., Ballhausen, R., et al. 2024, \apjl, 973, L25

\bibitem[Xrism Collaboration et al.(2025)]{cross} Xrism Collaboration, Audard, M., Awaki, H., et al.\ 2025, \aap, 702, A147. doi:10.1051/0004-6361/202556000

\bibitem[Yaqoob \& Murphy(2011)]{yaqoob11} Yaqoob, T. \& Murphy, K.~D.\ 2011, \mnras, 412, 277. doi:10.1111/j.1365-2966.2010.17908.x

\bibitem[Yaqoob et al.(2003)]{yaqoob2003} Yaqoob, T., McKernan, B., Kraemer, S.~B., et al.\ 2003, \apj, The Kinematics and Physical Conditions of the Ionized Gas in Markarian 509. I. Chandra High Energy Grating Spectroscopy, 582, 1, 105. doi:10.1086/344541

\bibitem[Yaqoob \& Padmanabhan(2004)]{yaqoob2004} Yaqoob, T. \& Padmanabhan, U.\ 2004, \apj, The Cores of the Fe K Lines in Seyfert 1 Galaxies Observed by the Chandra High Energy Grating, 604, 1, 63. doi:10.1086/381731

\bibitem[Yaqoob \& Serlemitsos(2005)]{yaqoob} Yaqoob, T. \& Serlemitsos, P.\ 2005, \apj, 623, 1, 112. doi:10.1086/428432

\bibitem[Yaqoob(2012)]{yaqoob12} Yaqoob, T.\ 2012, \mnras, 423, 3360. doi:10.1111/j.1365-2966.2012.21122.x

\bibitem[Zhao et al.(2021)]{zhao2021} Zhao, X., Civano, F., Fornasini, F.~M., et al.\ 2021, \mnras, 508, 4, 5176. doi:10.1093/mnras/stab2885





\end{thebibliography}
\end{document}